\documentclass[11pt,a4paper]{article}
 \pdfoutput=1

\usepackage{graphicx}
\usepackage{dcolumn}
\usepackage{bm}
\usepackage{hyperref}
\usepackage{empheq}
\usepackage{amsfonts}
\usepackage{amsthm}
\usepackage{graphicx}
\usepackage{hyperref}
\usepackage{soul} 
\usepackage{url}
\usepackage{color}
\usepackage{bm}
\usepackage{subfig}
\usepackage{bbm}
\usepackage{bigints}
\usepackage{authblk}
\usepackage{blindtext}

\newcommand{\ncd}{\newcommand}
\ncd{\mrm}    {\mathrm}
\ncd{\beq} {\begin{equation}}
  \ncd{\eeq} {\end{equation}}
  \def\d{{\rm d}}
  
\def\E{{E}}
\def\p{{\bm{p}}}
\def\M{{\bm{M}}}

\def\L{{\mathcal{L}}}
\def\d{{\mathrm{d}}}

\begin{document}

\title{From entropic to energetic barriers in glassy dynamics: 
The Barrat--M\'ezard trap model on sparse networks 
}

\author[1]{Diego Tapias \thanks{Corresponding Author: diego.tapias@theorie.physik.uni-goettingen.de}}
\author[2]{ Eva  Paprotzki}%
\author[3]{Peter Sollich}
\affil[1,2]{%
Institut f\"ur Theoretische Physik, University of G\"ottingen, Friedrich-Hund-Platz 1, 37077 G\"ottingen, Germany
}%
\affil[3]{%
Institut f\"ur Theoretische Physik, University of G\"ottingen, Friedrich-Hund-Platz 1, 37077 G\"ottingen, Germany
}%
\affil[3]{Department of Mathematics, King's College London, London WC2R 2LS, UK}

\date{\today}


\maketitle

\begin{abstract}
Trap models describe glassy dynamics as
a stochastic process on a network of configurations representing local energy minima.
We study within this class the paradigmatic Barrat--M\'ezard 
model, which has Glauber transition rates. Our focus is on the effects of the network \emph{connectivity}, where we go beyond the usual mean field (fully connected) approximation and consider sparse networks, specifically random regular graphs. 
We obtain the spectral density of relaxation rates of the master operator using the cavity method, revealing very rich behaviour as a function of network connectivity $c$ and temperature $T$. We trace this back to a crossover from initially entropic barriers, resulting from a paucity of downhill directions, to
energy barriers that govern the escape from local minima at long times.
The insights gained are used to rationalize the relaxation of the energy after a quench from high $T$, as well as the corresponding correlation and persistence functions. 
\end{abstract}

\section{Introduction}

Trap models have been widely used to model glassy dynamics~\cite{dyre1987master, bouchaud_weak_1992, bouchaud_aging_1995, barrat_phase_1995, monthus_models_1996, melin_glauber_1997, rinn_multiple_2000,  bertin_cross-over_2003, denny_trap_2003, doliwa_energy_2003, arous_dynamics_2006, sollich_trap_2006, moretti_complex_2011, baity-jesi_activated_2018, cammarota_numerical_2018, woillez2019active}.  They abstract the full dynamics of supercooled liquids, structural glasses and amorphous systems more generally, into motion near local potential energy minima (inherent structures), interspersed with relatively rare transitions between minima. Each minimum with its surrounding potential energy basin is identified as a discrete ``trap'', and the possible transitions define a stochastic dynamics on a network of these traps. In more sophisticated versions of the approach, traps are defined by clustering nearby basins with rapid transitions between them  into so-called meta-basins~\cite{denny_trap_2003, doliwa_energy_2003, yang2009dynamics}.
 
The most widely studied trap models are those in which the set of traps is fully connected, i.e.\ the mean field trap models. The full network connectivity allows these models to be studied analytically using different methods~\cite{bouchaud_weak_1992,  bouchaud_aging_1995, barrat_phase_1995, monthus_models_1996, melin_glauber_1997,  arous_dynamics_2006, arous_arcsine_2008, gayrard_aging_2010}, yielding predictions both for the relaxation of single-time quantities like the average energy and for two--time correlations and responses. Much less is known for the more realistic case where the degree of each node is small compared to the number of traps. Our work in this paper is designed to fill this gap, and indeed we will find that sparse network connectivity has very rich physical effects that have no analogue in mean-field trap models.

In addition to the {\em network connectivity} as a crucial ingredient in our analysis, a trap model is defined by the {\em energies} of the traps, and the {\em transition rates} between traps. These should satisfy detailed balance with respect to the trap energies, to ensure that the eventual equilibrium that a system of finite size would reach is a Boltzmann distribution. These aspects all remain fixed during the time evolution, so for theoretical analysis it is convenient to think of the network structure and trap energies -- which determine the transition rates -- as quenched disorder drawn from appropriate distributions~\cite{margiotta2018spectral,bovier2005spectral, akimoto_non-self-averaging_2018}. 

Arguably the simplest trap model is the Bouchaud model, which has Arrhenius transition rates that presume activation to some common threshold energy level and depend only on the departing trap.
For this model recent studies~\cite{margiotta2018spectral, margiotta2019glassy} have considered in detail the effect of the structure of the network of traps, with a focus on the spectral and time domain properties of the master operator that defines the dynamics. In this paper we use a similar set of tools to analyze the Barrat--M\'ezard (BM) trap model on Random Regular Graphs (RRG). The BM model uses Glauber transition rates, which are essentially constant for transitions that lower the energy but of Arrhenius form for energy increases. This is an attractive feature that allows the model to encode both conventional {\em energy} barriers that need to be surmounted by thermal activation and {\em entropic} barriers that arise when the number of possible downhill transitions from typical traps becomes small~\cite{bertin_cross-over_2003, cammarota2015spontaneous}.  Thus the model includes the two main mechanisms that are responsible for the slow dynamics in glasses. 

The spectral properties we concentrate on are the relaxation rates of the stochastic system, i.e.\ the negative eigenvalues of the master operator. This operator is a large random matrix due to the disorder in network structure and trap depths and so we import tools from random matrix theory for our analysis, specifically the Cavity Method~\cite{rogers2008cavity, mezard2009information, metz2010localization}.  
From the distribution of the relaxation rates one can predict qualitatively how the 
system will relax in the time domain as illustrated by recent studies for a range of different physical systems~\cite{amir2012relaxations, lahini_nonmonotonic_2017}.

In Sec.~\ref{model} we define the BM model, the RRG network structures we consider, and the associated master operator. The cavity method we use to work out its spectral properties is set out in Sec.~\ref{spectrala}. We present results (Sec.~\ref{results}) first for the mean-field limit of large connectivity, where we show that the relaxation rate spectrum is given by the distribution of the escape rates from the individual traps. This is followed by the analysis for the general case of sparse connectivity and  arbitrary temperature, with a focus on temperatures below the glass transition. In Sec.~\ref{timedomain} we translate the results for the relaxation rate spectrum  into the time domain to analyze the behaviour of a one-time observable -- specifically the mean energy -- and two-time correlation and persistence functions. We conclude in Sec.~\ref{conclu} with a discussion of the results and perspectives for future work.

\section{Barrat--M\'ezard trap model}
\label{model}

\subsection{Barrat--M\'ezard dynamics}

The Barrat--M\'ezard model considers dynamics on a set of traps that we label by $i=1,\ldots,N$. For traps $i$ and $j$ that are connected, i.e.\ are neighbours in the network of traps, the transition rate  
is of Glauber form, 
\beq
W_{ji} := W_{i \rightarrow j} = \frac{1}{c}\, \frac{1}{1 + \exp(-\beta (E_j - E_i))} \, ,
\label{dynamics}
\eeq
where the normalization by $c$, the average number of neighbours, ensures that the total escape rate from a typical trap is of order unity. 
As usual, $\beta=1/T$ is the inverse temperature ($k_{\rm B} = 1$). $E_i$ and $E_j$ are the depths, i.e.\ the negative energies, of the traps; we set the origin of the energy scale so that $E_i > 0$ for all $i$. Two limiting cases can be obtained from~\eqref{dynamics}: for $E_j-E_i \gg T$ , $W_{ji} \rightarrow 1/c$, which means that transitions to significantly deeper traps take place at a rate that is independent of  temperature. 
This is what allows entropic effects to occur in the dynamics, especially at low $T$ where the system goes downhill in the energy landscape
~\cite{barrat_phase_1995, bertin_cross-over_2003, cammarota2015spontaneous}. In the opposite limit  $E_i - E_j \gg T$ one has $W_{ji} \propto \exp(-\beta(E_i - E_j))$; this Arrhenius form is the signature of the thermally activated nature of transitions to higher energy traps~\cite{baity-jesi_activated_2018}. 

{Physically the model described by the transition rates~\eqref{dynamics} assumes that there are no barriers for any possible transition between configurations. This suggests that any trap with one or more neighbours lying lower in energy should be interpreted as a saddle rather than a local minimum of the energy landscape, and has led to the name ``step model'' being used in some of the literature, see e.g.~\cite{cammarota2015spontaneous, carbone2020effective}.
This interpretation is not unique, however, because one can equivalently read the model as having a nonzero uniform barrier for any transition.
To see this, first write the Glauber transition rate from $i$ to $j$ in terms of activation to an effective barrier state of depth $E_b$: }
\beq
{W_{i \to j} \propto \exp(-\beta (E_i -E_b))}
\eeq
{This maintains detailed balance if $E_b$ is a symmetric function of $(E_i,E_j)$; the standard Glauber rates~\eqref{dynamics} correspond to the choice }
\beq
E_b = -\frac{1}{\beta} \ln\left(\exp(-\beta E_i)+ \exp(-\beta E_j)\right)
\eeq
{If we now shift the barrier state upwards by a constant energy, $-E_b \to -E_b + \mathrm{const.}$, then $E_i-E_b$ is never smaller than this constant and so every transition has a finite energy barrier. The effect on the dynamics, however, is a simple (temperature dependent) rescaling of all transition rates so the physics remains the same as in the original model. This shows that it is equally valid to think of the states in the Barrat-M\'ezard model as genuine traps, i.e.\ local energy minima.}

\subsection{Energy distribution}
\label{edist}

The distribution of trap depths is usually taken to be either exponential or Gaussian. The latter choice leads to a REM-like model~\cite{cammarota_numerical_2018} in which the energy may be positive or negative. The exponential distribution, which was used in the original mean field Bouchaud trap model~\cite{bouchaud_weak_1992, monthus_models_1996} and much of the literature since, is more physical if we think of the traps as representing local energy minima {and it is justified from extreme--value statistics}~\cite{ bouchaud_universality_1997}. The threshold at depth $E=0$ 
can then be understood as the top of this energy landscape of local minima. In the following we choose the width of the trap depth distribution as our energy scale, which in dimensionless units is then just $\rho_E(E) =  \exp(- E)$. The Glauber transition rates of the BM model obey detailed balance with respect to the Boltzmann distribution {, as can be checked explicitly from~\eqref{dynamics}}, so if the system equilibrates then the probability of finding it in a trap of depth $E$ is $\propto \rho_E(E)e^{\beta E} = e^{(\beta-1)E}$. This distribution becomes unnormalizable at the temperature $T_{\rm g}=1$, which defines the glass transition. For $T<T_{\rm g}$ the system ages towards deeper and deeper traps and it is this regime of glassy dynamics that we will mostly focus on.

\subsection{Network structure}

To represent the potentially sparse connectivity of the network of traps we model this network as one of the simplest choices, a random regular graph (RRG) with connectivity $c$. This means explicitly that the network is randomly selected among all networks where every node (trap) has exactly $c$ neighbours. Consistent with the notation above, $c$ is then also the average number of neighbours. To ensure sparse connectivity we will keep $c$ finite while taking the thermodynamic limit $N\to\infty$, so that always $c\ll N$.

The RRG contains the essential features of sparse random networks by confining all dynamical transitions to the local environment of a node. As appropriate for a configuration space model it is also infinite-dimensional in the sense that the number of nodes grows exponentially with distance. In this work we focus on  connectivity $c \geq 3$. This condition ensures that the fraction of nodes outside the giant connected component vanishes for large $N$~\cite{margiotta2018spectral}, which  
implies that the network is connected as expected on physical grounds.

For our analysis it will be important that, because of their sparse connectivity, RRGs become locally tree-like for $N\to\infty$.  This is a consequence of the fact that typical loop lengths are $\sim \ln(N)$ and so diverge in the limit~\cite{albert_statistical_2002}. The locally tree-like structure is what enables us to use the cavity method~\cite{rogers2008cavity, mezard2009information} to obtain the spectral properties, in a way {that becomes exact for $N\to\infty$. }

We mention in passing that an RRG is effectively a Bethe lattice with no boundaries~\cite{biroli2018delocalization} and thus different from a Cayley tree. This technical detail is relevant in the context of localization of eigenvectors on Bethe lattices~\cite{de2014anderson, sonner2017multifractality, biroli2018delocalization, tikhonov2019statistics}.

\subsection{Master equation}

We can now write down the master equation governing the time evolution of the vector $\p(t) = (p_1(t), \ldots, p_N(t))$ of probabilities of being in any of the $N$ traps. If $A_{ij}$ is the adjacency matrix of the network of traps, with $A_{ij}=1$ if two traps are connected and $=0$ otherwise, then the master equation reads
      \beq
      \frac{\d \p(t)}{\d t} = {\M} \p(t) \, .
      \label{mastere}
      \eeq
The master operator  $\M$ has elements
      \beq
      M_{ji} = {A_{ji}} W_{i \rightarrow j}  \, , \qquad M_{ii} = -\sum_{j \neq i} M_{ji}
      \eeq
As can be seen from their definition, the (negative) diagonal are the  rates of escape $-M_{ii}=\Gamma_i$ from the individual traps.
We recall that the transition rates $W_{i \rightarrow j}$ are given by~\eqref{dynamics} and the trap depths are sampled independently from the exponential distribution $\rho_E$ defined in Sec.~\ref{edist}.
    
\section{Spectral Analysis}
    \label{spectrala}

The general solution of the master equation~\eqref{mastere} can be written as~\cite{margiotta2018spectral}
  \beq
\p(t) = \sum_{\alpha = 0}^{N-1} {\rm{e}}^{\lambda_\alpha t} ( {\bm{L}}_\alpha, \p(0)) {\bm{R}}_\alpha \, ,
\label{eq:general_soln}
\eeq
where the ${\bm{L}}_\alpha$ and ${\bm{R}}_\alpha$ are the left and right eigenvectors of $\M$, respectively, and the $\lambda_\alpha$ are the corresponding eigenvalues. These are non-positive and we order them in the following so that $0 = \lambda_0 \leq -\lambda_1 \leq -\lambda_2 \leq \ldots \leq -\lambda_{N-1}$; note that the connectedness of the network ensures that there is only a single zero eigenvalue, which describes equilibrium~\cite{anderson1991continuous}.

Eq.~(\ref{eq:general_soln}) shows that the dynamics is a superposition of exponentials with relaxation rates $-\lambda_\alpha$. The spectrum of these relaxation rates therefore play a key role in the physical behaviour. It is described in the thermodynamic limit by the eigenvalue distribution or spectral density
\beq
\rho(\lambda) = \lim_{N\rightarrow \infty} \frac{1}{N} \left \langle \sum_{\alpha=0}^{N-1} \delta(\lambda - \lambda_\alpha) \right \rangle \, ,
\label{spectral}
\eeq
The average here is over the graph realizations (structural disorder) and the local trap depths (energetic disorder), though for large $N$ this is essentially equivalent to considering a single realization because of self-averaging.

\subsection{Methods}

In order to develop a theory for the spectrum we first symmetrize the master operator. Because of detailed balance, this can be achieved using a similarity transformation with the equilibrium distribution,
\beq
\M^s = {\bm P}_{\rm{eq}}^{-1/2} \M   \,  {\bm P}_{\rm{eq}}^{1/2}
\label{ss}
\eeq
Here ${\bm P}_{\rm{eq}}$ is a diagonal matrix with entries $({\bm P}_{\rm{eq}})_{ii} = p_i^{\rm{eq}}$ and the $p_i^{\rm{eq}} \propto e^{\beta E_i}$ are the equilibrium occupations of the traps. Such a similarity transformation preserves the spectrum, but the fact that $\M^s$ is symmetric is important for setting up the cavity theory
~\cite{margiotta2018spectral}. 

In the same spirit we write the transition rates~\eqref{dynamics} as effective Bouchaud rates, which are of Arrhenius form, times a symmetric function:
 \beq
  W_{ji} =  \frac{{\rm{e}}^{\beta(E_i + E_j)/2}}{2\cosh(\beta(E_i - E_j)/2)} \frac{{\rm{e}}^{-\beta E_i}}{c} =: K(E_i, E_j)  \frac{{\rm{e}}^{-\beta E_i}}{c} \, .
  \label{fbeta_def}
 \eeq
 
The spectral density can in general be deduced from the properties of the resolvent ${\bm G}$ matrix, which is defined as~\cite{kuhn_spectra_2008, rogers2008cavity, livan2018introduction}
 \beq
 {\bm G}(\lambda - i \epsilon) = \left( (\lambda - i \epsilon) {\bm I} - \M^s \right)^{-1} \, ,
 \label{resolv}
 \eeq
by the relation
\begin{equation}
    \rho^{\M}(\lambda) = \lim_{\epsilon \rightarrow 0} \frac{1}{\pi N} \sum_{j=1}^N {\rm{Im}} \, G_{jj} \, ,
    \label{specific}
  \end{equation}
 {where in the equation~\eqref{resolv}, ${\bm I}$ denotes the identity matrix.} As indicated by the superscript, the spectrum here is initially for a fixed master operator and finite $N$, while $\epsilon$ is a regularizer that broadens the delta-distributions in the definition of the spectrum into Lorentzians of width $\epsilon$.

The cavity method starts from the fact that the resolvent entries $G_{jj}$ can be thought of as the variances of an $N$-dimensional Gaussian distribution with exponent $-\bm{x}^{\rm T} \bm{G}^{-1}\bm{x}/2$.
This exponent has nonzero cross terms $x_i x_j$ only for neighbouring traps $i$ and $j$, so inherits the structure of the network of traps. If the network is a tree, the variance of each $x_j$ can then be worked out by what is known in the machine learning literature as Belief Propagation~\cite{mezard2009information}: the ``marginal'' variance $1/\omega_j$ at any node can be expressed in terms of the ``cavity variances'' $1/\omega_k^{(j)}$ of the neighbours.  ``Cavity'' here means that $1/\omega_k^{(j)}$ is the variance at node $k$ if node $j$ had been removed from the network, i.e.\ a cavity created; $\omega_k^{(j)}$ can therefore also be called a cavity precision or cavity Green's functions~\cite{biroli2018delocalization}. Intuitively, $\omega_k^{(j)}$ is a ``message'' that node $k$ sends to node $j$ and that encodes all properties of the part of the network that feeds into $j$ via $k$.

In practice the equations take a simpler form if one rescales the variables $\bm{x}$ in the Gaussian probability distribution so that the resolvent elements are expressed in terms of the 
marginal variances as
\beq
G_{jj} = i\frac{{\rm{e}}^{\beta E_j} c}{\omega_j
} 
\eeq
The relation between $\omega_j$ and the cavity precisions is then 
 \beq
\omega_j = i (\lambda - i \epsilon) {\rm{e}}^{\beta E_j}c +  \sum_{k \in \partial j} \frac{i K(E_j, E_k) \omega_k^{(j)} }{i K(E_j, E_k)  + \omega_k^{(j)}  } \, ,
\label{marginals}
\eeq
where $\partial j$ indicates the set of neighbours of $j$.
The cavity precisions themselves can be obtained from a set of equations that is almost identical except that node $j$ is removed from the sum: 
\beq
\omega_k^{(j)} = i (\lambda - i \epsilon) {\rm{e}}^{\beta E_k}c  + \sum_{l \in \partial k \setminus j} \frac{i K(E_k, E_l) \omega_{l}^{(k)} }{i K(E_k, E_l) + \omega_l^{(k)} } \, .
\label{cavities}
 \eeq
 Once these cavity equations have been solved, the spectrum  for a given master operator $\M$ can be obtained from equation~\eqref{specific} with~\eqref{marginals} inserted, i.e.
 \beq
 \rho^\M(\lambda) = \lim_{\epsilon \rightarrow 0} \frac{1}{\pi N} \sum_{j = 1}^N {\rm{Re}} \left( {\rm{e}}^{\beta E_j} c/\omega_j \right) \, .
 \label{rhosingle}
 \eeq
 In line with the fact that the function $K$ is the factor by which BM transition rates differ from those in the Bouchaud model (see~\eqref{fbeta_def} ), Eqs.~(\ref{marginals}, \ref{cavities}) reduce to those for the Bouchaud case~\cite{margiotta2018spectral} if we replace $K$ by $1$.

The cavity equations~\eqref{cavities} are in principle approximate on any finite random regular graph because of the presence of loops, but in the thermodynamic limit $N\to\infty$ they are expected to become exact as loop lengths diverge. In that limit one also sees that, for given $E_k$, the sum on the r.h.s.\ of (\ref{cavities}) consists of $c-1$ statistical independent terms, each relating to the traps in only one branch of the (tree-like) network. One can therefore reduce the description to a {\em distribution} $\zeta(\omega,E)$ of cavity precisions. We emphasize that as each $\omega_k^{(j)}\equiv \omega$ depends on the energy $E_k \equiv E$ of the {\em sending trap}, such a joint distribution is required, in contrast to the simpler Bouchaud case~\cite{margiotta2018spectral}. The cavity equations~(\ref{cavities}) then become a self-consistent equation for $\zeta$:
 \begin{align}
  { \zeta(\omega, E) = \rho_E(E) \int \prod_{ j=1}^{c-1} \d E_j \d \omega_j \delta( \omega - \Omega_{c-1})\zeta(\{\omega_j, E_j\}) } \, ,
 \label{selfc}
 \end{align}
with the abbreviation 
 \begin{align}
{ \Omega_{a}(\{\omega_l, E_l\}, E) = i\lambda_\epsilon {\rm{e}}^{\beta E} c +  \sum_{l=1}^{a} \frac{i K(E, E_l) \omega_{l}}{i K(E, E_l) + \omega_l} }
\label{omegas}
 \end{align}
 and $\lambda_\epsilon \equiv \lambda - i \epsilon$.
 Eq.~\eqref{selfc} can then be solved using a Population Dynamics algorithm~\cite{abou1973selfconsistent, biroli2018delocalization}.  In this method as adapted to our case one starts from a finite population of $(\omega,E)$-pairs, initialized from some in principle arbitrary initial distribution. The population is then updated iteratively by picking $c-1$ random members  
$(\omega_l, E_l)$ 
and a new energy $E$ from the exponential distribution $\rho_E(\cdot)$, calculating $\Omega_{c-1}(\{\omega_l,E_l\},E)$ and replacing a random population member by $(\Omega_{c-1},E)$~\footnote{Numerical code implemented in Julia can be found at~\url{https://github.com/dapias/SparseBarratMezard}}. After the population has equilibrated, we then calculate the spectral density~\eqref{specific} from the population analogue of (\ref{rhosingle}), as
 \beq
 \rho(\lambda) = \lim_{\epsilon \rightarrow 0} \frac{1}{\pi} {\rm{Re}} \left \langle \frac{{\rm{e}}^{\beta E} c}{\Omega_c(\{\omega_l, E_l\}, E)} 
 \right \rangle_{(\{\omega_l, E_l \}, E) } \, .
 \label{rholambda}
 \eeq
{For all the results presented below, a population of $10^4$ pairs $\{ (\omega, E) \}$ was used unless otherwise specified. {For every given $\lambda$, $10^8$ iterations were performed to reach equilibrium, with $\epsilon$ set to $\epsilon_0 = 10^{-300}$. 
 The population was then evolved for a further $10^7$ steps and at each update step a sample ($\Omega_c, E$) was drawn for the calculation of the average~\eqref{rholambda}.
 We evaluate  this average simultaneously for different values of $\epsilon$, usually $\{10^{-3}, 10^{-4} \}$, to assess the $\epsilon$-dependence of the results. For a more detailed discussion of the role of $\epsilon$ we refer to~\cite{margiotta2018spectral}. }}
 
\section{Results: Spectral properties}
\label{results}
\subsection{Escape rates versus relaxation rates}
\label{MF}
We first consider the mean-field case $c\to\infty$, where the relaxation rate spectrum has previously been worked out for the Bouchaud~\cite{bovier2005spectral} but not the BM trap model~{(with the exception of~\cite{melin_glauber_1997} where the case $T =0$ is analyzed).} We next show that in this limit the relaxation rates become the (negative) rates $\Gamma_i= \sum_{j \neq i} W_{ji}$,  of escape from individual traps, 
with finite $c$-corrections scaling as $1/c$:
 \beq
\lambda_i = -\Gamma_i + O(1/c)
\label{lambda_correction}
\eeq
To see this, note from Eqs.~(\ref{ss},\ref{fbeta_def}) that one can decompose the symmetrized master operator $\M^s$ as
\beq
\M^s = \M^{(0)} + \frac{1}{c}\M^{(1)}
\label{pert}
\eeq
where $\M^{(0)}$ is a diagonal matrix with elements
\beq
\M^{(0)}_{ij} = - \frac{\delta_{ij}}{c} \sum_{k \neq i} \frac{A_{ki}}{1 + {\rm{e}}^{-\beta(E_k - E_i)}} = -\delta_{ij} \Gamma_i \, ,
\label{nzmatrix}
\eeq
which are nothing other than the negative escape rates. The off-diagonal terms are collected in  $\M^{(1)}$, which has elements (recall that  $A_{ij} = 0$ if $i = j$)
\beq
\M_{ij}^{(1)} = \frac{A_{ij}}{2 \cosh \left( \beta (E_i - E_j)/2 \right) } \, .
\eeq
We now exploit the fact that the escape rates in the ``baseline'' operator $\M^{(0)}$ remain of order unity for $c\to\infty$, while the contribution from $\M^{(1)}$ in \eqref{pert} scales with $1/c$. Using standard perturbation theory~\cite{melin1997glauber, griffiths2018introduction}, the eigenvalues of $\M^s$ can then be expanded as
\beq
\lambda_i = -\Gamma_i + \frac{1}{c}\lambda_i^{(1)} + \frac{1}{c^2} \lambda_i^{(2)} + \ldots \, ,
\eeq
with
\beq
\lambda_i^{(1)} = \M_{ii}^{(1)} = 0\, , \qquad  \lambda_i^{(2)} = \sum_{j \neq i} \frac{(\M_{ij}^{(1)})^2}{ -\Gamma_i + \Gamma_j} \, ,
\eeq
The sum defining $\lambda_i^{(2)}$ has $c$ entries whose size is independent of $c$, hence $\lambda_i^{(2)}=O(c)$ and we get the $1/c$ scaling of the corrections announced in \eqref{lambda_correction}. This implies in  particular that $\lambda_i \to -\Gamma_i$ in the mean field limit $c \rightarrow \infty$, so that the escape rate distribution directly determines the spectrum of relaxation rates.

The above discussion implies that there is in fact a second limit where relaxation rates and escape rates become identical, namely $T\to 0$ at arbitrary $c$. This is simply because in that limit $\beta\to\infty$ and so $\M^{(1)}\to 0$. We will see below that the distribution of (relaxation or escape) rates then degenerates into a sum of $c+1$ delta peaks.

\subsection{Escape rate distribution}
Given the result of the previous subsection, we next consider the distribution of escape rates. For the sake of generality we do this for arbitrary $c$ to start with. The escape rate distribution can then formally be written as
\beq
\rho_\Gamma(\Gamma) = \langle \delta(\Gamma - \hat{\Gamma}) \rangle_{\{E, E_1, \ldots, E_c\}}
\label{distgamma}
\eeq
with
\beq
\hat{\Gamma}{(\{E_1, \ldots, E_c\}, E)} = \frac{1}{c} \sum_{j=1}^c \frac{1}{1+ {\rm{e}}^{-\beta(E_j - E)}} \, ,
\label{gamma}
\eeq
The average in~\eqref{distgamma} is over the exponential distribution from which each of the $c+1$ trap depths is independently drawn.
The distribution $\rho(\Gamma)$ can then be constructed numerically simply by the appropriate sampling. Analytically, the distribution can be obtained explicitly for the (unphysical) case $c=1$, with the result (see details in Appendix~\ref{escc1}):
\begin{align}
\rho_\Gamma(\Gamma) = \frac{T}{2} \left\{
\begin{array}{lll}
\Gamma^{T-1}
(1-\Gamma)^{-T-1}
&\mbox{for}&
\Gamma<1/2\\
\Gamma^{-T-1}
(1-\Gamma)^{T-1}
&\mbox{for}&
\Gamma>1/2
\end{array}
\right.
             \,.
               \label{escdist}
\end{align}
This expression shows that for $\Gamma \rightarrow 0$, the escape rate density diverges as $\Gamma^{T-1}$. 
As demonstrated in section~\ref{twotime} below (cf.\ Eq.~\eqref{re2}), this divergence in fact controls the small $\Gamma$-behaviour of $\rho_\Gamma(\Gamma)$ for {\em any} finite $c$. 
%
%
Remarkably,
the exponent $T-1$  is identical to the one found in the spectrum of relaxation rates in the Bouchaud model~\cite{margiotta2018spectral}, which is characteristic of activated processes with an exponential distribution of barrier heights. We thus have here the first signature of the fact that the BM model with finite $c$ has activated features when we look at low rates, i.e.\ deep traps.

\begin{figure}
\centering
  \includegraphics[width = \textwidth]{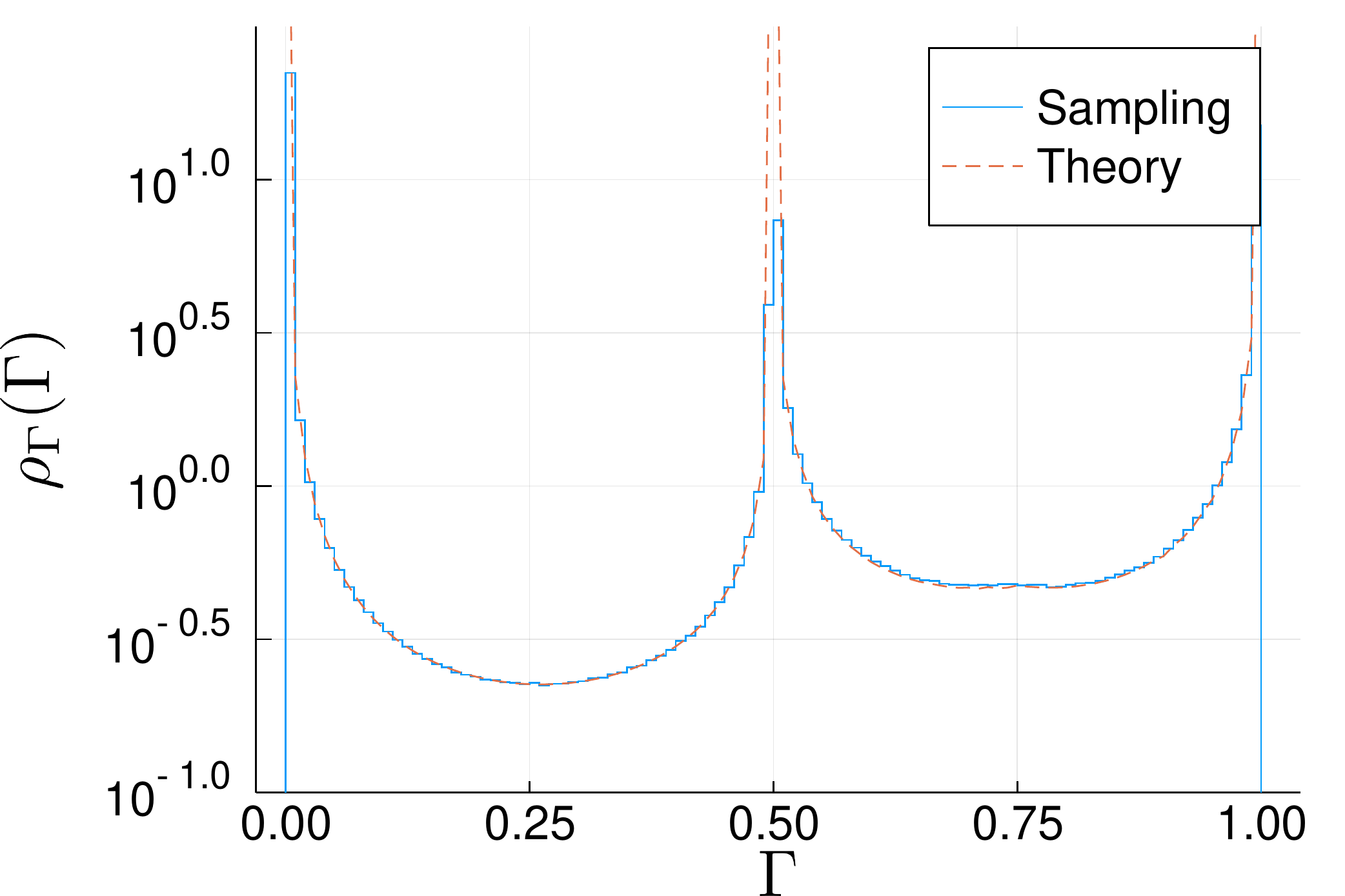}
  \caption{Escape rate distribution obtained by analytical calculation (see Appendix~\ref{escc1}) and direct evaluation by numerical sampling of~\eqref{distgamma} for $c=2$ and $T=0.1$.
  }
  \label{fig:escdist}
\end{figure}

By way of preparation for the discussion of the relaxation rate spectra, we mention that for $c\geq 2$ the escape rate distribution can have singularities not just at $\Gamma=0$ and 1 but at any multiple of $1/c$ between these two extreme values~{(see also Appendix~\ref{largec}).}
For the case $c=2$ one finds, for example, that the relaxation rate density at $\Gamma=1/2$ is controlled by the integral
(see details in Appendix~\ref{escc1})
\beq
\rho_\Gamma(\Gamma = 1/2) \sim \int_0^{1/2} \frac{(1 - R_1)^{-3T-2}}{R_1^{-3T + 2}} \d R_1 \, ,
\label{cequal2}
\eeq
and thus divergent for $ T < 1/3$. This gives some intuitive justification for the fact that similar divergences at $-\lambda = i/c$, $i = 1, \ldots, c-1$ will be found below in the relaxation rate spectrum.
The escape rate distribution for $c=2$ and $T=0.1$ is shown in Figure~\ref{fig:escdist} and displays the expected singularities at $\Gamma=0,1/2$ and 1. The plot also demonstrates agreement between the analytical evaluation of~\eqref{distgamma} and direct sampling of~\eqref{gamma}.

We motivated our discussion of the escape rate distribution by the fact that it also gives the relaxation rate spectrum for $c\to\infty$; we defer a quantitative evaluation of this mean field limit to the next section. In the other case where the two distributions coincide, i.e.\ $T\to 0$, they are straightforward to work out. The escape rates from~\eqref{nzmatrix} simplify to
\beq 
\Gamma_i = \frac{1}{c}\sum_{k\neq i} A_{ki}\Theta(E_k-E_i)
\eeq 
where $\Theta(\cdot)$ is the Heaviside step function. Each $\Gamma_i$ is then just the fraction of deeper ($E_k>E_i$) traps among the neighbours of traps $i$. This fraction can take the values $i/c$ with $i \in \{ 0, 1, \ldots, c \}$. If we order the $c+1$ depths of a given trap and its $c$ neighbours into a descending list, then a relaxation rate of $i/c$ results precisely when the central trap is at position $i+1$ in this list, because it then has $i$ lower-lying (deeper) neighbours. But by permutation symmetry the central trap is equally likely to be in {\em any} position in the sorted list of $c+1$ trap depths so that
\beq
\rho_\Gamma(\Gamma) = 
\frac{1}{c+1} \sum_{i = 0}^{c} \delta \left(\Gamma - {i}/{c} \right) 
\label{esct0}
\eeq 
The above argument for the uniform prefactor $1/(c+1)$ of the delta peaks can be confirmed by explicit calculation as sketched in Appendix~\ref{prefa}. 

\subsection{Relaxation rate distribution}

We now turn to the relaxation rate spectrum, which encodes information about collective relaxation modes. In contrast to the escape rate distribution, it therefore cannot generally be found from just local information. The spectrum depends on two key parameters, temperature $T$ and network connectivity $c$. We analyse their effects separately, beginning with the former.
 
\begin{figure}
  {{\includegraphics[width = \textwidth]{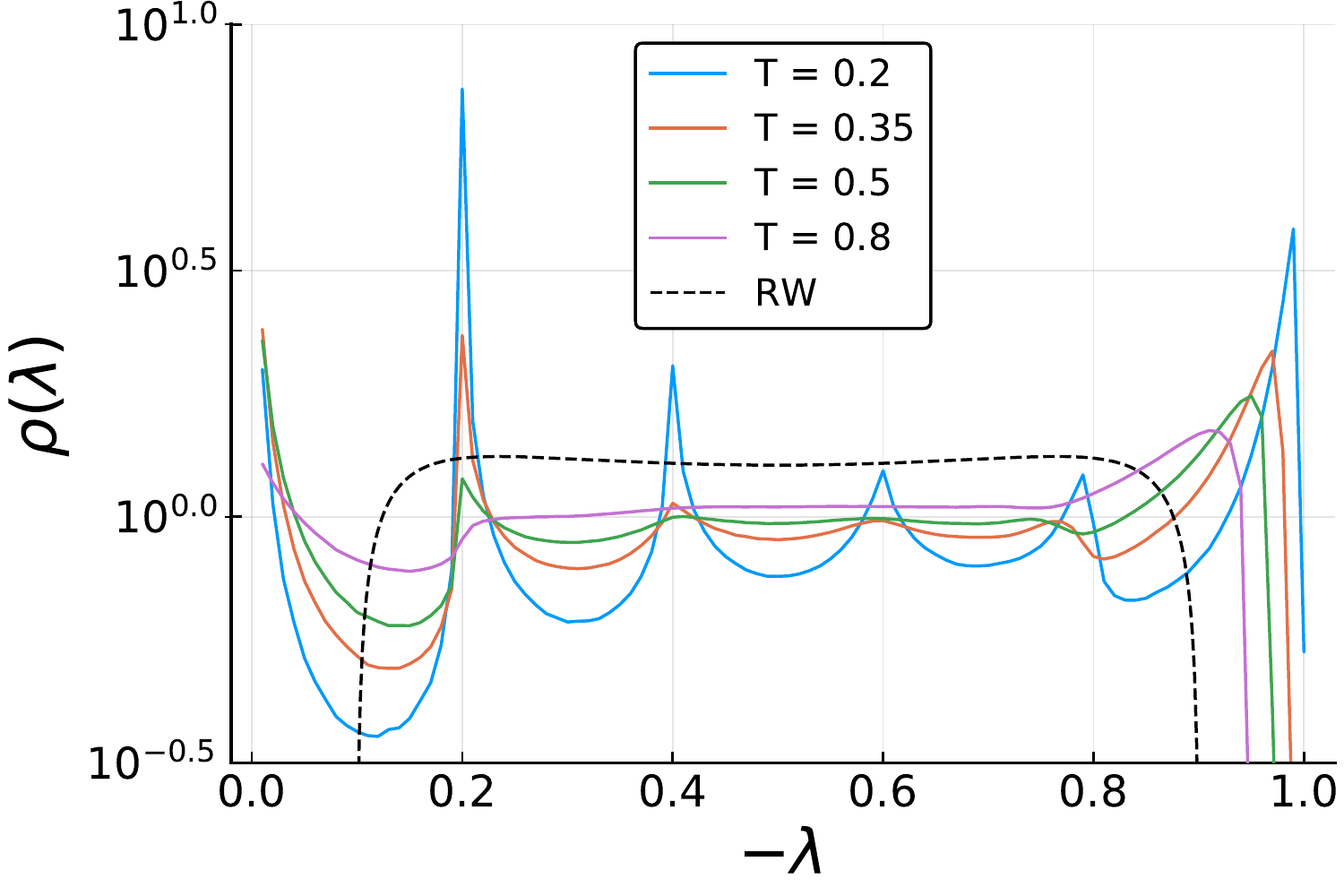} }}
     \caption{$T$-dependence of the spectral density for $c =5$, evaluated from the cavity theory (see Eq.~\eqref{rholambda}) with smoothing parameter $\epsilon = 10^{-3}$. Also shown is the limiting behaviour for $T\to\infty$ as given by the random walk spectrum (Kesten--McKay law~\eqref{kesten}).}
          \label{tchanges}
        \end{figure}

\begin{figure}
    \includegraphics[width =  \textwidth]{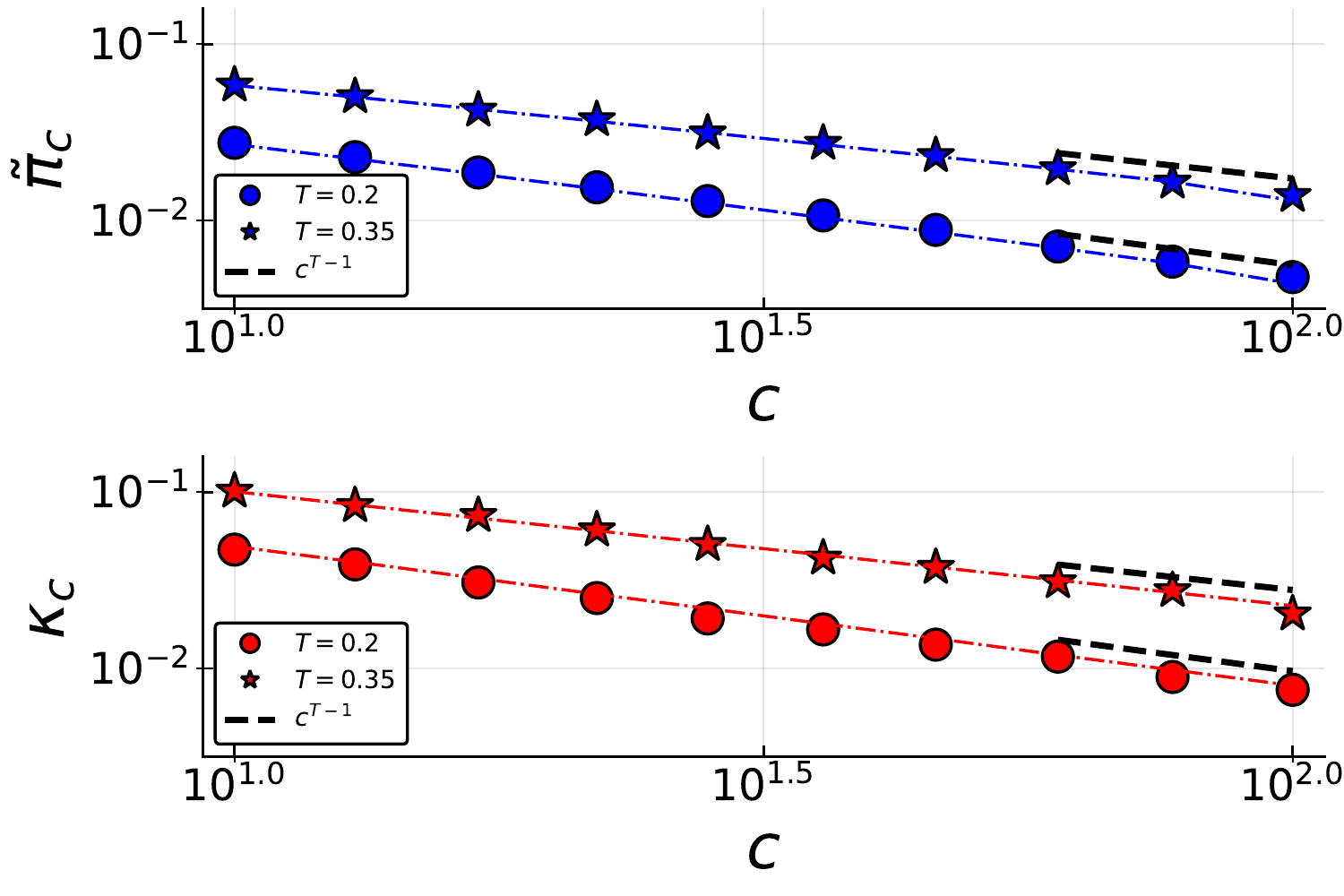}
    \caption{Log-log plot of prefactors of the spectra in the small rate regime against connectivity $c$, for two different temperatures $T$. Top: Escape rates. The markers are obtained by numerical sampling of~\eqref{distgamma} in the range $\Gamma=10^{-10}$ to $10^{-6}$. The dot-dashed lines show the theoretically predicted  prefactor~\eqref{tilde_pi_c}.
    The dashed lines show the asymptotic power law scaling~\eqref{re2} with $c$. Bottom: Relaxation rates. The markers are obtained from numerical evaluation of the spectrum at $-\lambda = 10^{-9}$ for $T=0.2$ and $-\lambda = 10^{-7}$ for $T=0.35$. The dot-dashed lines show the theoretically predicted prefactor~\eqref{small2}, evaluated as explained in
    Appendix~\ref{cavlarge}. 
    The dashed lines show the asymptotic power law scaling~\eqref{small} with $c$.
    }
     \label{powerlawcs}    
   \end{figure}

\emph{Changes in $T$.} Using the cavity method presented in section~\ref{spectrala} we can obtain the variation of the spectral density with temperature for given $c$ in the sparse network regime ($c<\infty$). Exemplary results for $c=5$ are shown in Figure~\ref{tchanges}. They display a rich structure, with multiple peaks in the spectrum at low $T$, but as we now demonstrate the qualitative behaviour can be understood from the two extreme cases of high and low $T$. For $T \rightarrow \infty$, the trap depths are irrelevant as the Glauber transition rate between any two connected traps energy becomes $1/(2c)$. The dynamics thus becomes a random walk and the spectral density depends only on the network structure. For large random regular graphs the resulting spectral density $\rho(\lambda)$ is given explicitly by the Kesten-McKay law for infinite regular trees~\cite{kesten1959symmetric, chung1997spectral, mckay1981expected, bauerschmidt2019local}, which with our choice of transition rates reads
\beq
{\rho^{\rm{RW}} (\lambda) = \frac{2c}{\pi (1-4(\lambda + 1/2)^2)} \sqrt{\frac{c-1}{c^2} - \left(\lambda+ \frac{1}{2} \right)^2}}
\label{kesten}
\eeq
and is also plotted in Figure~\ref{tchanges}. For $T\to 0$, on the other hand, the escape rate analysis presented in the previous section shows that the relaxation rate spectral density consists of a sum of delta peaks of equal height at $-\lambda = i/c$, cf.\ Eq.~\eqref{esct0}. Qualitatively, the results in Figure~\ref{tchanges} can therefore be understood as interpolating between the relatively flat Kesten-McKay spectrum at high $T$ and a series of peaks at low $T$. An obvious question then concerns the {\em shape} of these peaks at low but nonzero $T$; we next show that they are power law divergences, with exponents that depend on $T$ and on the position of the peak as we already saw for
the escape rate distribution. 
We begin with the peak at $\lambda=0$, which corresponds to the distribution of the slowest modes that govern the aging dynamics at $T<1$. We find that the spectral density in this region grows as 
\begin{align}
  \label{small2}
  \rho(\lambda) &\simeq 
  \kappa_c |\lambda|^{T-1}  
  \end{align}
with a prefactor that for large $c$ scales as
  \begin{align}
\kappa_c \sim c^{T-1} 
\label{small}
\end{align}
%
{so that overall $\rho(\lambda)\sim |c\lambda|^{T-1}$ for large $c$ and small $\lambda$. Specifically one requires $\lambda\ll 1/c$ to see this power law scaling, as one can show by an analysis similar to the one in Appendix~\ref{largec}.} {The scaling~\eqref{small2} may be obtained analytically from the cavity equations as shown in  Appendix~\ref{cavlarge}. This includes the prefactor $\kappa_c$ and its scaling (\ref{small}) for large $c$. 
Figure~\ref{powerlawcs} shows the evaluation of this prefactor across a range of $c$, and clearly confirms the theoretical predictions.
The scaling of $\rho(\lambda)$ with $\lambda$, though not the prefactor, can also be derived within a Single Defect Approximation as explained in Appendix~\ref{sda}. 
} 

\begin{figure}
    {{\includegraphics[width =  \textwidth]{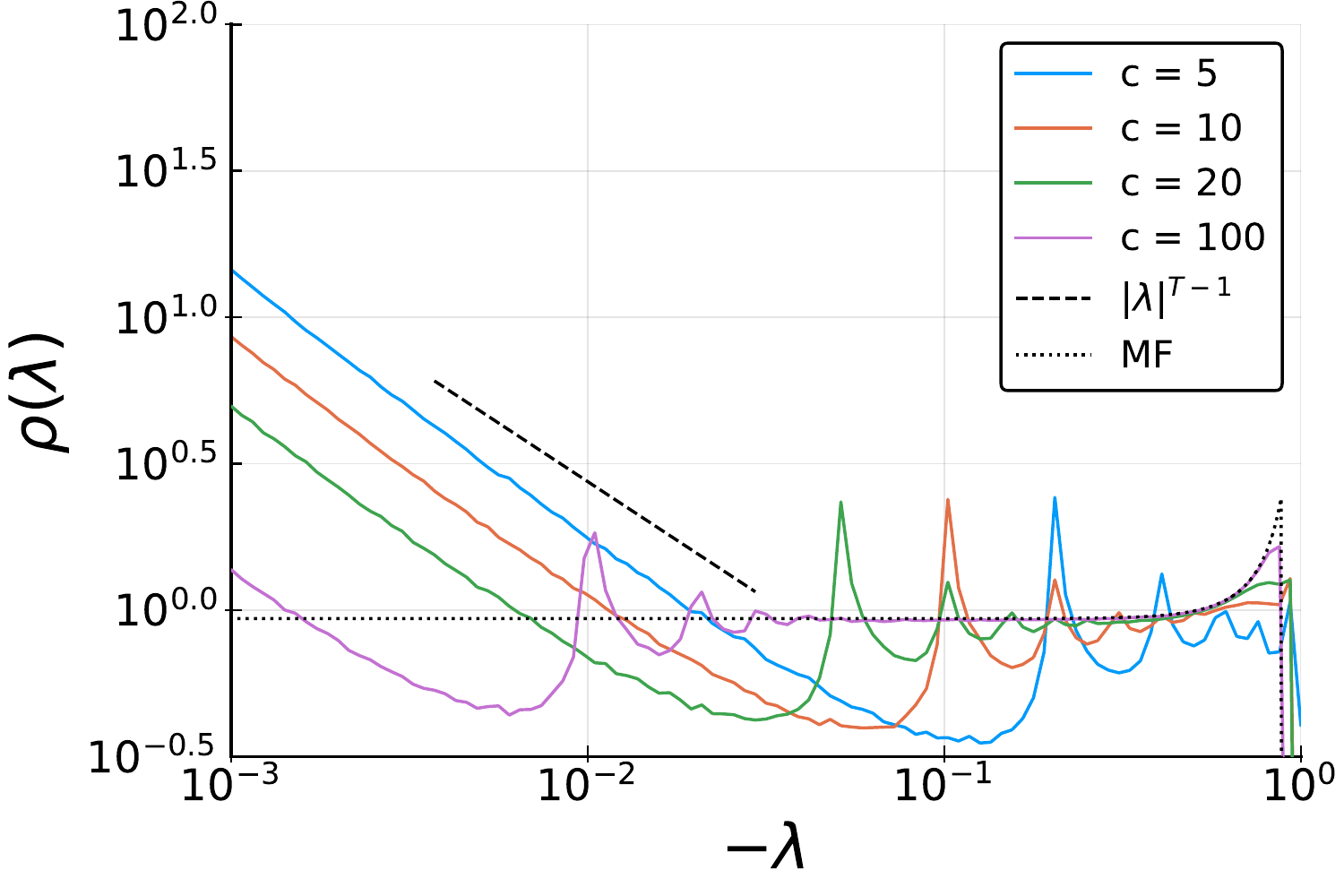} }}
    \caption{Log-log plot of spectral density for $T = 0.2$ and different values of $c$, as obtained from the cavity method with $\epsilon = 10^{-4}$. The power law asymptote for small $|\lambda|$ is consistent with the theoretical prediction $\sim |\lambda|^{T-1}$ (dashed line)~\eqref{small2}. The mean field limit which can be obtained from the numerical inversion of~\eqref{mfesc} is also shown (dotted line). The plateau  for small $|\lambda|$ is predicted by~\eqref{MF_plateau}.}
     \label{cchangeswsda}
   \end{figure}

To summarize thus far, for any finite $c$ the power law dependence of $\rho(\lambda)$ for small $\lambda$ is the same as in the Bouchaud trap model on a RRG~\cite{margiotta2018spectral}. We thus conclude that also the collective relaxation modes are governed by activated processes at long times. The physical picture is that the dynamics of the system is then dominated by slow transitions between local energy minima, i.e.\ traps that have no ``escape directions'' in the form of lower-lying neighbours. We validate the predicted $\lambda$-scaling in Figure~\ref{cchangeswsda}, for fixed $T$ and different $c$.

\begin{figure}
  \includegraphics[width= \textwidth]{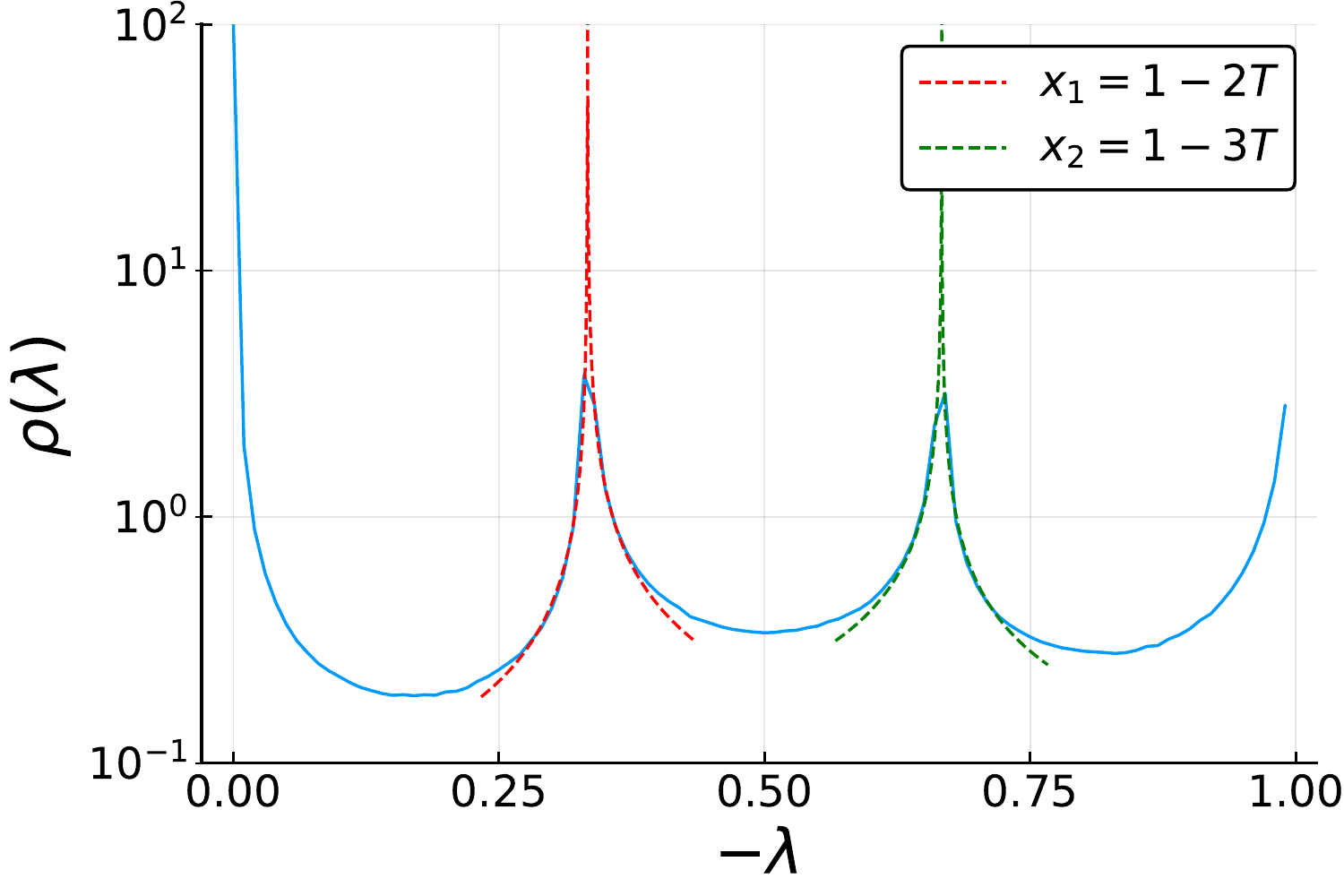}
  \caption{Spectral density for $c = 3$ and $T = 0.1$ obtained via the cavity method with $\epsilon = 10^{-3}$. At $\lambda_i^* = -i/c$ the spectrum diverges as a power law with the exponents indicated in the legend.}
 \label{diverg}
    \end{figure}

The low temperature behaviour of the other spectral peaks, around $\lambda^*_i= -i/c$ with $i = \{1, \ldots, c\}$, is more intricate. Numerical evaluation using the cavity theory suggests that all peaks (except the one near $\lambda^*_c=-1$) are power law singularities up to some limiting temperature, so that the spectrum around them takes the form $\rho(\lambda) \sim |\lambda + i/c|^{-x_i}$. The exponent $x_i$ depends on the peak $i$ being considered and in general also on $c$. It decreases with temperature $T$; where it drops to zero the corresponding peak turns into a maximum of finite height. The temperatures where this occurs all lie in the range $0<T\leq 1/2$.
In Figure~\ref{diverg}, the spectrum for $c=3$ at low temperature is displayed together with the power law fit for the intermediate peaks. For the first peak at $\lambda_1^*=-1/c$, our numerical data for the exponent $x_1$ across a range of temperatures are consistent with $x_1=1-2T$ independently of $c$, while the exponents of higher peaks ($x_2$ etc.) do have a nontrivial $c$-dependence.

The existence of intermediate peaks in the spectral density is characteristic of the Barrat--M\'ezard trap model and has no analogue in the Bouchaud model~\cite{margiotta2018spectral}. { These peaks reflect the network structure and hence entropic effects; on the other hand their {\em broadening} results from activation effects. They can be seen as an extension to nonzero $T$ of the delta peaks at $T=0$, where the dynamics is driven purely by entropic effects, and then must be interpreted as a signature of the entropic barriers that are present in the model and dominate the dynamics for $T<1/2$}~\cite{bertin_cross-over_2003, cammarota2015spontaneous}. 

{Comparing now the qualitative features of the spectral density and the escape rate spectrum, we have seen that they agree exactly at $T=0$ and remain qualitatively similar at low $T$. With increasing $T$ differences appear, e.g.\ the power law divergences at nonzero $\lambda$ or $\Gamma$ have different exponents and disappear at distinct temperatures in the two types of spectra. The divergences of the spectra for the slowest modes (small $-\lambda$ or $\Gamma$) survive up to $T=1$, on the other hand, and have the same exponent. These observations have implications for the localization properties of the relaxation modes: escape rates are local quantities so their spectrum is expected to agree with that of the relaxation rates only where the relaxation modes are also localized on a small number of nodes. Our results thus suggest that at low $T\ll 1$ much of the relaxation rate spectrum is {\em localized}, while at higher $T$ most relaxation modes except for the slowest ones are extended. A quantitative analysis of these $T$-dependent localization properties will be presented in a separate work.
}

\emph{Changes in $c$.} Having so far focused on the dependence of the spectral density on $T$, we now keep $T$ fixed and consider the variation with $c$. As an example, Figure~\ref{cchangeswsda} shows results for $T=0.2$. For small $c$ we observe spectral peaks as before. These become more numerous with increasing $c$ but also less pronounced, merging for very large $c$ into a smooth spectrum.  This mean-field limit can be obtained explicitly; to our knowledge even this is a new result. We exploit the discussion in section~\ref{MF}, which showed that for $c\to\infty$ the relaxation rate distribution is identical to the escape rate distribution. The latter can be obtained by noting that for large $c$ the escape rate~\eqref{gamma} becomes a deterministic function of the trap depth: 
   \begin{align}
     \Gamma(E) &= \lim_{c\rightarrow \infty} \frac{1}{c} \sum_{j =1}^c \frac{1}{1+{\rm{e}}^{-\beta(E_j - E)}} \\
               &= \int_0^\infty \d E'  \, \frac{\rho_{E}(E') }{1+{\rm{e}}^{-\beta(E' - E)}} \label{Gamma_E_integral}
               \\
               &=   \, _2F_1(1, T, 1+T,-{\rm{e}}^{\beta E}) \, ,
                 \label{mfesc}
   \end{align}
with $_2F_1$ the Gauss hypergeometric function. From this relation the escape rate distribution $\rho(\Gamma)$ is then found by a simple variable transformation, $\rho(\Gamma) = \rho_E(E(\Gamma)) \lvert \frac{\d E}{\d \Gamma} \rvert$. In general the inverse function $E(\Gamma)$ has to be found numerically. But for large $E$, i.e.\ deep traps with low escape rates, the lower limit of the integral~\eqref{Gamma_E_integral} can be sent to $-\infty$ with negligible error provided we are in the glass phase, $T<1$, giving (see also~\cite{sollich_trap_2006})
 \beq
 \Gamma(E) 
 \approx \frac{\pi T}{\sin(\pi T)} {\rm{e}}^{-E} \, ,
 \label{mfescape}
\eeq
As this is proportional to $\rho_E(E)$, it results in a {\em  constant} escape rate density and hence also relaxation rate density,
\beq
\rho(\lambda) = \frac{\pi T}{\sin(\pi T)} \, , \qquad {\lambda \rightarrow 0}.
\label{MF_plateau}
\eeq
in other words the spectrum is flat for small $\lambda$. 
Of course as we saw above, for any finite $c$ the spectral density eventually has to cross over
to the divergence~\eqref{small2} for small $\lambda$. Comparing with~\eqref{MF_plateau} shows that the crossover point must scale as $-\lambda \sim 1/c$ and so moves towards $\lambda=0$ as the mean field limit of large $c$ is approached. {This physically means that the connectivity of the network sets the scale for which the crossover from entropic to energetic dynamics is observed (more on this in the next section).}
These features can be seen qualitatively by looking back at Figure~\ref{cchangeswsda}. 

 \section{Results: Time domain properties}
 \label{timedomain}
\subsection{Numerical simulation approach}

  
For time domain properties we obtain stochastic simulation results directly for the thermodynamic limit $N\to\infty$, by generating an effectively infinite tree on the fly, during the course of a stochastic simulation using the Gillespie algorithm. The network construction method is identical to that used for the Bouchaud trap model in~\cite{margiotta2019glassy}; its key advantage is that it allows us to study time-dependent properties without any finite size effects. The Gillespie simulation algorithm consists of repeated evaluation the following steps:
\begin{enumerate}
 \item For a given node $i$ compute the total exit rate $\Gamma_i = \sum_{j\in \partial i} W_{ji}$.
\item Compute the waiting time $\Delta t>0$ until the next transition to a neighbouring node by sampling from  $p_i(\Delta t) = \Gamma_i \exp(-\Gamma_i  \Delta t)$. 
\item Select the node $i^{\rm{new}}$ to which the transition occurs  randomly among the $c$ neighbours of $i$, with probability $W_{ji}/\Gamma_i$ for node $j$.
\item Return $i^{\rm{new}}$ as the new node and increment the running simulation time by $\Delta t$. 
 \end{enumerate}
  
We store the nodes visited as a function of time for many different realizations. This allows us to see the evolution of the mean energy $\langle E(t) \rangle$ and also obtain two time quantities, namely the correlation  $C(0, t)$ and the persistence $P(0, t)$, that we can compare with results from the relaxation rate and escape rate spectra. All of these quantities are evaluated with respect to the initial state for the dynamics; we assume that this is a uniform distribution across traps, corresponding to an equilibrium distribution at large temperature ($T\to \infty$).


\subsection{ Mean energy}
\label{meanE}
   \begin{figure}
    {{\includegraphics[width =  \textwidth]{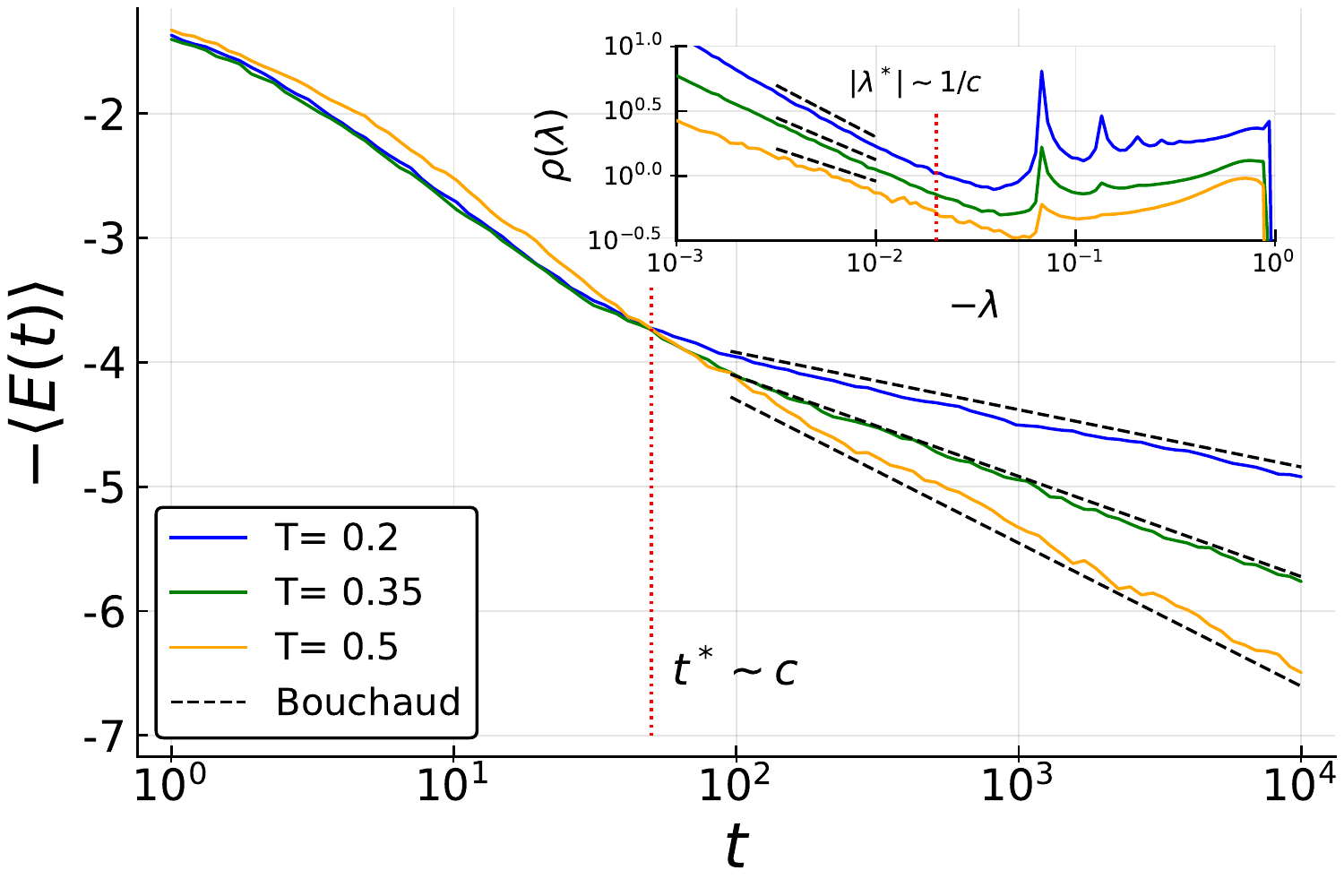}}}
    \caption{Mean energy as a function of time for $c = 15$, averaged over $10^4$ different trajectories. Inset: spectrum of relaxation rates for the same temperatures. The vertical line in both figures indicates the existence of a crossover regime beyond which (for large $t$, or small $|\lambda|$) activated behavior is found. 
    }
    \label{meanecconstant}
  \end{figure}
  Probably the simplest manifestation of the aging dynamics of trap models in the glass regime is the fact that the average energy decreases continually with time; in this sense the system keeps track of its age through its energy~\cite{bertin2008laser}. In Figure~\ref{meanecconstant} we show what form this energy decay takes in the BM model with finite connectivity $c$, for three different temperatures. Beyond an initial transient the mean energy decay is logarithmic in time, but with a clear change in slope at some crossover time. From the spectral point of view, this crossover time $t^*$ can be understood in terms of a crossover relaxation rate $|\lambda^*| \sim  1/t^*$ below which the spectral density  $\rho(\lambda) $ follows the activated behavior~\eqref{small2}. This crossover is indicated in the inset of Figure~\ref{meanecconstant}. 

\begin{figure}
   {{\includegraphics[width =  \textwidth]{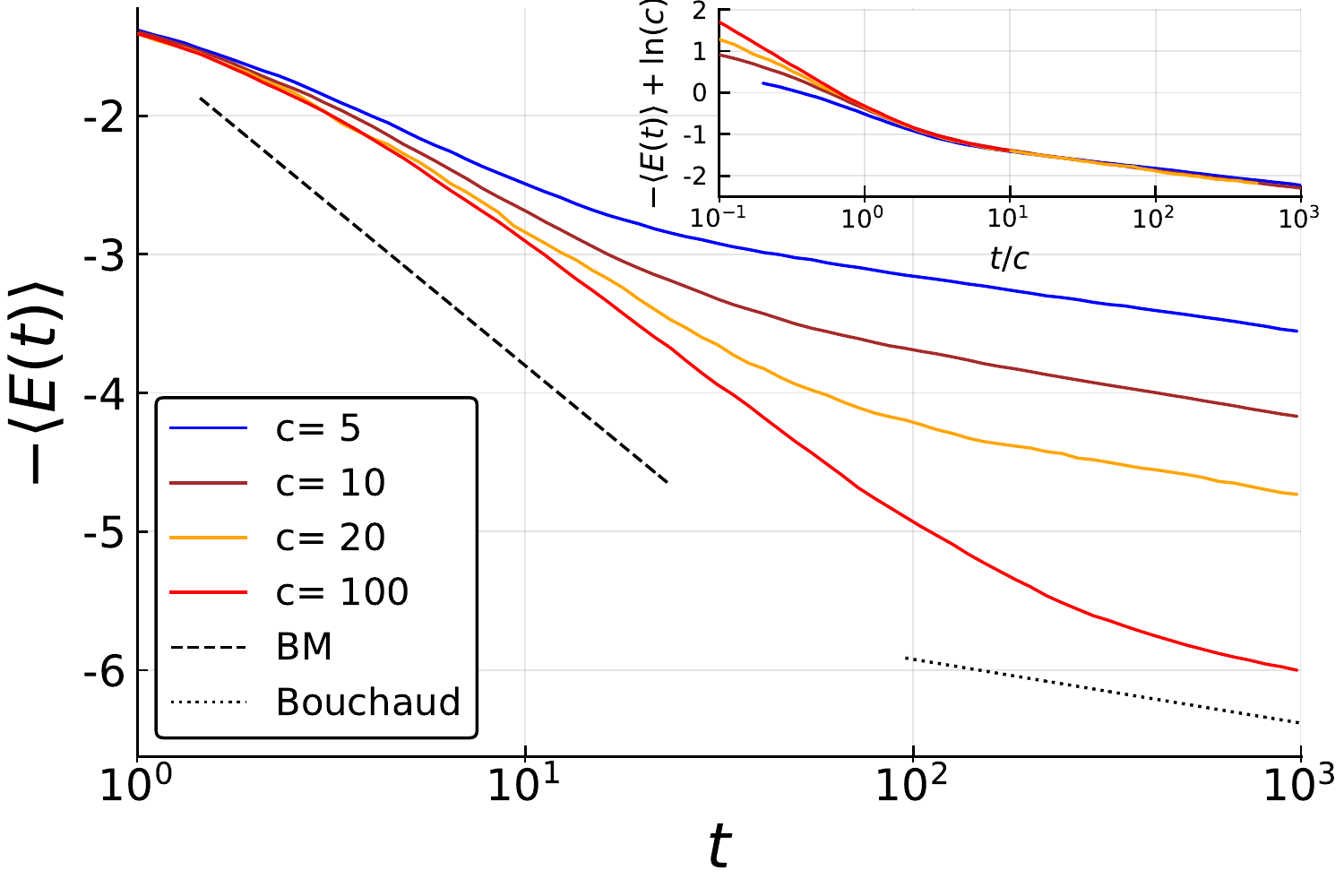}}}
    \caption{Mean energy as a function of time for $T = 0.2$, averaged over $10^4$ different trajectories. The dashed lines correspond to the behavior expected from the mean field Bouchaud model and the mean field Barrat-- M\'ezard (BM) model. Inset: Collapse to a master curve for large $c$ after rescaling by the crossover time $t^ * \sim c$, see text for details.}
     \label{meanetconstant} 
   \end{figure}
The notion of a change in the dynamics between an entropic, non--activated and an activated regime is also apparent from Figure~\ref{meanetconstant}, where the dependence on $c$ of the energy decay is displayed together with the predicted behavior from the mean-field Bouchaud and Barrat--M\'ezard models, which is~\cite{monthus_models_1996, bertin_cross-over_2003}
\begin{align}
  \langle E(t) \rangle \sim
  \begin{cases}
    - T \ln(t) \qquad {\mbox{Bouchaud}} , \\
    -\ln(t) \qquad {\mbox{Barrat M\'ezard}} \, .
  \end{cases}
  \label{energy_decay}
\end{align}
It is clear that the crossover between these two regimes shifts towards longer times as $c$ increases. This of course has to be so, as for $c\to\infty$ the mean-field BM behaviour must be obtained for any finite $t$. 
In quantitative terms, it follows by comparison of the small $\lambda$ divergence ~\eqref{small2} of the spectrum with the mean-field plateau~\eqref{MF_plateau} that the crossover relaxation rate $\lambda^*$ has to scale with $1/c$ and therefore $t^* \sim  c$. We confirm this in the inset of Figure~\ref{meanetconstant}, where we show the energy decay plotted against $t/c \sim t/t^*$. Applying the corresponding shift $-E \to -E + \ln c$ to the energy axis, which accounts for the entropic (BM) relaxation up to $t\approx t^*$ (see  Eq.~\eqref{energy_decay}), shows a rather good collapse of the curves for different values of the connectivity $c$.
   
\subsection{Two-time observables}
\label{twotime}

\begin{figure}
  \centering
          {{\includegraphics[width =  \textwidth]{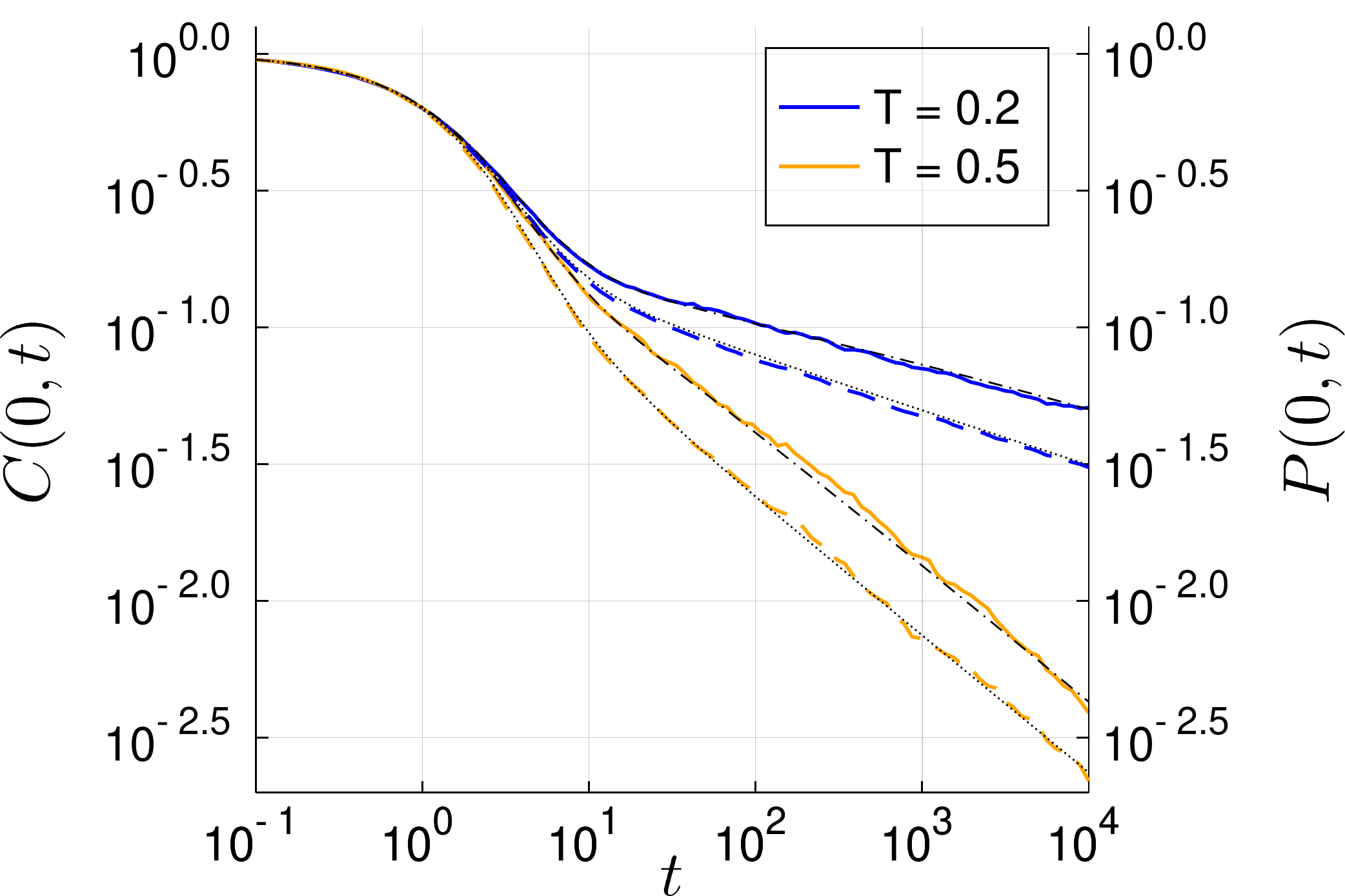}}}
\caption{Correlation and persistence function for $c = 5$ and two different $T$. Colored lines: Stochastic simulation results averaged over $10^4$ trajectories; the persistence is the lower curve in each pair of curves. Dash-dotted lines: Cavity theory predictions for correlation using~\eqref{lapcavi}. {Dotted lines:} Analytical prediction for persistence~\eqref{persinum}.}
     \label{correlplot}
   \end{figure}
   
 In this final section we study the correlation $C(0, t)$ and persistence $P (0, t)$ (or survival probability) as functions of the observation time $t$; the initial time will always be fixed at ``waiting time'' 
 $t_{\rm w} = 0$. The correlation is defined as the probability of finding the system in the same trap at time $0$ and $t$. If the system starts in trap $i$, then from the master equation~\eqref{mastere} and its solution~\eqref{eq:general_soln} this probability is the $ii$-element of the propagator, $(e^{\M t})_{ii}$. Averaging over the uniform initial distribution across traps yields
   \beq
   C(0, t) =  
   \frac{1}{N} \sum_{i=1}^N 
   ({\rm{e}}^{\M t})_{ii}
   = \frac{1}{N}{\rm Tr\,}{\rm{e}}^{\M t} 
   = \frac{1}{N} \sum_{\alpha=0}^{N-1}  {\rm{e}}^{\lambda_\alpha t}   \, ,
   \label{cc}
   \eeq
 In the limit of large $N$, Eq.~\eqref{cc} becomes
   \beq
   C(0, t) = \int \d \lambda \, \rho(\lambda)  {\mathrm{e}}^{\mathbf{\lambda} t}
   \label{lapc}
   \eeq
so that the correlator can be deduced from the spectral density that we can already predict using the cavity method. An essentially equivalent connection can be made to the trace of the resolvent: from~\eqref{lapc}, the Laplace transform of the correlator is 
\beq 
\L[C(0,t)] = \frac{1}{N}\sum_{\alpha=0}^{N-1} \frac{1}{s-\lambda_\alpha} = \frac{1}{N}{\rm Tr\,}(s{\bm I}-\M)^{-1}
\eeq
As the eigenvalues of the master operator $\M$ and its symmetrized version $\M^s$ are identical, the last expression is the normalized trace 
of the resolvent~\eqref{resolv}, giving
\begin{align}
  C(0,t) &= \L^{-1} \left[\frac{1}{N} \sum_{i=1}^N {G_{ii}}(s) \right] 
  ,         \label{lapcavi}
\end{align}
The resolvent trace can again be obtained using the cavity method and the correlator obtained by numerical inverse Laplace transform. This is numerically more convenient than  finding the spectrum first and then apply~\eqref{lapc}. Figure~\ref{correlplot} compares the results with averages over $10^4$ stochastic simulation trajectories. The agreement is excellent,
providing an explicit demonstration of how spectral information can be used to obtain non-trivial temporal properties of the BM trap model dynamics.

We turn next to the persistence function $P(0,t)$. This is defined as the probability of not leaving the initial trap up to time $t$. If the initial trap is $i$, this probability can be expressed in terms of the escape rate $\Gamma_i$ of the trap as $\mathrm{e}^{-\Gamma_i t}$. Averaging again over the initial traps gives
\begin{align}
  P(0,t) &=
  \frac{1}{N} \sum_{i=1}^N {\mathrm{e}}^{-\Gamma_i t} 
  = \int \d \Gamma \rho_\Gamma(\Gamma)\mathrm{e}^{-\Gamma t}
  \label{persis_rho_Gamma}
  \, ,
  \end{align}
where the last expression applies in the limit of large $N$ and relates the persistence to the  escape rate distribution, by direct analogy with~\eqref{lapc}. Inserting the expressions~(\ref{distgamma},\ref{gamma}) for this distribution gives explicitly
\begin{align}
 &P(0, t) 
=  \langle {\rm{e}}^{- (t/c) \sum_{j=1}^c 1/[1 + \exp(-\beta(E_j- E))]} \rangle_{E, E_1, \ldots, E_c} \\
   &= \int \d E\,  \rho_E(E)  \left[\int \d E_1 \, \rho_E(E_1) {\rm{e}}^{-\frac{t}{c}/[1 + \exp(-\beta(E_1 - E) ]} \right]^c  .
      \label{persinum}
\end{align}
This integral can be evaluated numerically and the result 
compared to that calculated by averaging over stochastic simulation trajectories, as shown in Figure~\ref{correlplot} alongside the correlation results. As it should be the agreement is very good. The long-time scaling of the persistence can also be worked out analytically from the integral~\eqref{persinum} as discussed in Appendix~\ref{perscal}, giving 
\begin{align}
P(0, t) &\simeq \pi_c t^{-T} 
\label{persistence_scaling}
\end{align}
with $\pi_c\sim c^{T-1}$ for large $c$. 
By computing the inverse Laplace transform in~\eqref{persis_rho_Gamma} we obtain that
\begin{align}
  \rho_\Gamma(\Gamma) & \simeq  \tilde{\pi}_c \Gamma^{T-1}  
\label{re2}
\end{align}
where (with $\Gamma(\cdot)$ the Euler Gamma function)
\beq 
\tilde{\pi}_c = \pi_c/\Gamma(T)
\label{tilde_pi_c}
\eeq 
{The comparison with~\eqref{small2} reveals that, for a given temperature, the spectra for the slowest processes (either of escape or relaxation) have the same scaling. This is clearly supported by the numerical data in Fig.~\ref{powerlawcs}.} We remark that the correspondence between $\rho(\lambda)$ and $\rho_\Gamma(\Gamma)$ translates in the time domain into a proportionality between $C(0,t)$ and $P(0,t)$ at long times, which again agrees with our numerics (see Fig.~\ref{correlplot}). 

 \section{Conclusions}
 \label{conclu}
 
We have focussed in this study on the characterization of the dynamics of the Barrat--M\'ezard trap model, which is a simple coarse-grained model for the  dynamics of glassy systems in configuration space. Its main feature is the combination of activated and non--activated pathways on an energy landscape consisting of traps with an exponential distribution of depths. Key for the importance of activated processes is the sparse connectivity of the network of traps in configuration space; we modelled this as a random regular graph. This is more general (and realistic) than the original Barrat--M\'ezard model, which is of mean-field type because its network is fully connected~\cite{barrat_phase_1995}.

Mathematically the model is a continuous time Markov chain on a random graph with microscopic transition rates given by the Glauber form~\eqref{dynamics}. The aim of this work was to obtain the spectral density of the associated master operator, in the thermodynamic limit. This was done via the Cavity Method. Our main finding is that below the glass transition temperature the slowest relaxation modes, which determine the long time behavior, have a density that follows a divergent power law (Eq.~\eqref{small2}), which is characteristic of activated processes~\cite{margiotta2018spectral, margiotta2019glassy}. This contrasts with the mean field case of infinite connectivity ($c\to\infty$), where the spectral density is constant for the slowest modes. We established this by showing that in this limit the relaxation rates reduce to local escape rates. This allowed us to calculate the spectrum exactly for the mean-field BM trap model, which to the best of our knowledge is also a new result. At low temperatures and finite connectivity we found power law singularities also for intermediate relaxation rates. We interpreted these as the finite temperature analog of the delta peaks that make up the spectrum for $T=0$. These delta peaks indicate that relaxation rates -- again equal to escape rates for $T\to 0$ -- are governed by the fraction of lower-lying neighbours of any given trap.

Comparing the eventual small $\lambda$-divergence ~\eqref{small2} of the spectral density for finite $c$ with the mean-field plateau we obtained a crossover relaxation rate $|\lambda^*| \sim 1/c$. 
Physically, this means that the network structure generates a crossover between {\em entropically dominated} dynamics at $t<1/|\lambda^*| \sim c$ to primarily {\em activated dynamics} for longer times. Numerical simulations data for the decay of the average energy, which we obtained via a bespoke algorithm that eliminates all corrections due to finite network size,   confirmed this behaviour.

The correlation and persistence functions following a quench from infinite temperature were also analyzed; we showed that these can be predicted from the spectrum of relaxation and escape rates, respectively, in excellent agreement with numerical simulation data.

In studying the BM trap model on networks with sparse connectivity, our broader aim was to make trap models into more accurate representations of real glassy systems. Indeed, the fact that the standard trap models have a mean field character has been criticized in a number of studies, and more realistic descriptions have been demanded~\cite{yang2009dynamics, moretti_complex_2011}. Our work is designed to fill this gap. What is fascinating is the natural appearance of a crossover in the glass regime between slow dynamics that are initially governed by entropy barriers but later become dominated by activation across energy barriers, directly from the sparse network connectivity. In particular the crossover emerges without the imposition of a restricted network size or of changes in the dynamical rules with trap depth~\cite{bertin_cross-over_2003}. Interestingly the entropic-energetic crossover does not appear in Bouchaud trap models on sparse networks~\cite{margiotta2018spectral, margiotta2019glassy}, presumably because the activated form of the local transition rates there rules out any downhill motion between traps for which entropic barriers would be relevant. {Our results are also relevant to what is still an ``open problem in glassy relaxation", namely the precise balance between activation and entropic relaxation~\cite{arceri2020glasses, stariolo2020barriers}. A comparison of the dynamics in finite-sized $p$-spin glasses with the mean field Bouchaud and Barrat--M\'ezard  predictions has been performed to study this question~\cite{ stariolo2020barriers, stariolo2019activated}; our results suggest that these insights could be deepened by considering the sparse version of those trap models. Such an approach could also provide an alternative point of view of the crossover from ``smooth''~\cite{stariolo2019activated} to activated dynamics in other finite-sized models with disorder (e.g.\ the Random Orthogonal Model)~\cite{crisanti2000potential, crisanti2000activated}.}

We comment briefly on the scaling of $c$ in connection to real-space glass physics. The number $N$ of local energy minima and hence traps is expected to scale exponentially in system volume $V$. The typical number of configurations reachable from a trap $i$, on the other hand, will scale as $V$ because transitions between traps will correspond to roughly independent local particle rearrangements inside finite volume elements {(which one could identify with cooperatively rearranging regions~\cite{adam1965temperature, stevenson2006shapes})}. The overall rate of transitions must then also scale as $V \sim c$. Our $O(1)$ transition rates would have to be scaled up by a factor $c$ to accommodate this, and timescales scaled down by $1/c$ accordingly. The entropic-energetic crossover time $t^*$ then becomes independent of $c$, so does {\em not} diverge with system size. This suggests that it is likely to be observable experimentally or in targeted simulations. 

Open questions for future work concern in particular the aging dynamics of the BM trap model on sparse networks. While for Bouchaud trap models the aging behaviour of two-time correlation functions $C(t_{\rm w},t)$ ultimately appears to become independent of the connectivity~\cite{arous_dynamics_2006, arous_arcsine_2008, ricthesis}, one would not expect this for the BM model where the crossover between entropic and energetic barriers introduces a separate timescale. The comparison to the aging scalings resulting from other approaches for generating such crossover timescales~\cite{bertin_cross-over_2003} should be particularly revealing.
The wider context of trap models as dynamical systems in a disordered potential also opens up further directions by making contact with different models in the area of disordered systems. The most relevant for our purposes is the Anderson model, in particular on sparse networks where it is the subject of much ongoing research~\cite{ tikhonov2016anderson, biroli2018delocalization}. A full understanding of Anderson localization in this case remains open and the connection with trap models may lead to new results, for instance, regarding dynamical universality classes.

\bibliographystyle{unsrt}
\bibliography{cavi}

\appendix

\section{Escape rate distribution}
\label{escc1}

The escape rate distribution~\eqref{distgamma} is most straightforward to analyze from its Fourier transform $\langle \mathrm{e}^{is\Gamma}\rangle$. As~\eqref{persis_rho_Gamma} shows, this is the same as the persistence function $P(0,t)$ with $t$ replaced by $-is$. Writing the exponential energy distributions in~\eqref{persinum} explicitly thus gives
\beq 
\langle \mathrm{e}^{is\Gamma}\rangle = 
 \int \d E\,  \mathrm{e}^{-E}  [\Delta(E,s/c)]^c
 \label{rho_Gamma_Fourier}
 \eeq
 with
 \beq 
 \Delta(E,s/c) = 
 \int \d E_1 \, \mathrm{e}^{-E_1} {\rm{e}}^{{i(s/c)}/{[1 + \exp(-\beta(E_1 - E) ]}}
 \eeq
If we make the change of variable
\begin{align}
R = \frac{1}{1 + {\rm{e}}^{-\beta(E_1 - E)}} \, ,
\end{align}
then $\d R = \beta R (1-R) \d E_1$ and $\mathrm{e}^{-(E_1-E)}=[(1-R)/R]^T$, yielding
\beq 
\Delta(E,s/c) = T\mathrm{e}^{-E}\int_{R_{\rm min}}^1 \d R\,\frac{(1-R)^{T-1}}{R^{T+1}}
\mathrm{e}^{i(s/c)R}
\eeq
The lower integration limit $R_{\rm min}=(1+\mathrm{e}^{\beta E})^{-1}$ lies between 0 and $1/2$, so for further calculation it can be helpful to split the integral into $\Delta = \Delta_+ + \Delta_-$, where $\Delta_+$ is the integral for $R=1/2\ldots 1$ and $\Delta_-$ the remainder. 

Now we focus on the case $c = 1$, where from~\eqref{rho_Gamma_Fourier} we just need the averages of $\Delta_\pm(E,s)$ over the exponentially distributed central trap depth $E$. The first of these is straightforward:
\begin{align}
  &\int_{0}^\infty {\d E} \, {\rm{e}}^{-E}  \Delta_+(E, s) = \notag \\
  & =T\int_0^\infty \d E \, {\rm{e}}^{-2E}     \int_{1/2}^{1}  \d R \,  {\rm{e}}^{i s R}\, \frac{(1-R)^{T-1}}{R^{T+1}} \\
  &= \frac{T}{2}   \int_{1/2}^{1}  \d R \,  {\rm{e}}^{i s R}\, \frac{(1-R)^{T-1}}{R^{T+1}}  \, .
  \label{delta_plus}
\end{align}
For the average of $\Delta_-$ we note that the lower integration limit $R_{\rm min}=(1+\mathrm{e}^{\beta E})^{-1}$ on $R$ corresponds to a lower limit of $E_{\rm min}=T\ln[(1-R)/R]$ for $E$ at fixed $R$:
\begin{align}
&\int_{0}^\infty {\d E} \, {\rm{e}}^{-E}  \Delta_-(E, s) = \notag \\
&=T\int_0^\infty \d E \, {\rm{e}}^{-2E}     \int_{R_{\rm min}}^{1/2}  \d R \,  {\rm{e}}^{i s R}  \frac{(1-R)^{T-1}}{R^{T+1}} 
\\
&= T     \int_{0}^{1/2}  \d R \,  {\rm{e}}^{i s R}\, \frac{(1-R)^{T-1}}{R^{T+1}} \,\frac{{\rm{e}}^{-2E_{\rm min}}}{2}
\\
&=  \frac{T}{2}    \int_{0}^{1/2}  \d R \,  {\rm{e}}^{i s R}\, \frac{(1-R)^{T-1}}{R^{T+1}} \left( \frac{1-R}{R} \right)^{-2T} \\
&=  \frac{T}{2}    \int_{0}^{1/2}  \d R \,  {\rm{e}}^{i s R}\, \frac{(1-R)^{-T-1}}{R^{-T+1}} \, .
\label{delta_minus}
\end{align}
Altogether we have now $\langle \mathrm{e}^{is\Gamma}\rangle =$~\eqref{delta_plus}$+$~\eqref{delta_minus}. The factors ${\rm{e}}^{i s R}$ in both terms just produce Dirac deltas $\delta(\Gamma-R)$ upon inverse Fourier transform, giving directly the result~\eqref{escdist}. The symmetry of the distribution under $\Gamma\to 1-\Gamma$ follows intuitively from the fact that the escape rate $R$ is transformed to $1-R$ when $E_1-E$ changes sign, together with the fact that this trap depth difference has an even distribution.

{For the case $c = 2$, the result in equation~\eqref{cequal2} is derived by computing $(\Delta_+ + \Delta_-)^2$, inverse Fourier transforming it and evaluating the remaining integrals for $\Gamma = 1/2$. The complete escape rate distribution is stated below as a piecewise function. For $ \Gamma \leq 1/2$ it is}
  \begin{align}
    \rho_{\Gamma}(\Gamma) &= \frac{4T^2}{3}  \int_0^{\Gamma} \d x \,  f(2\Gamma - x) g(x) \, ,
  \end{align}
{whereas for $ 1/2 < \Gamma \leq 3/4$,}
  \begin{align}
    \rho_{\Gamma}(\Gamma) &= \frac{4T^2}{3} \bigg( \int_{2\Gamma -1}^{1/2} \d x \,  f(2\Gamma - x) g(x)  + \notag \\ &+ \int_{1/2}^{{2\Gamma -1/2}} \d x  \dfrac{f(x) f(2\Gamma - x)}{2}\bigg) \, ,
  \end{align}
and finally for the remaining interval $ 3/4 < \Gamma \leq 1$:
  \begin{align}
    \rho_{\Gamma}(\Gamma) &=   \frac{2T^2}{3}  \int_{2\Gamma - 1}^{1}  \d x \, \dfrac{f(x) f(2\Gamma - x)}{2}  \,  ,
  \end{align}
with  $f(x) = \dfrac{(1 - x)^{T - 1}}{x^{T + 1}}$ and $g(x) = \dfrac{(1 - x)^{-2 T - 1}}{x^{-2 T + 1}}$. The above expression produces the plot in Fig.~\ref{fig:escdist}.

\section{Spectra for $T = 0$}
\label{prefa}

We used a permutation symmetry argument in the main text to justify why in the $T=0$ spectrum~\eqref{esct0} each delta peak has the same prefactor. This can also be seen explicitly as follows. The prefactor $a_k$ for 
each $\delta(\Gamma - k/c)$ in equation~\eqref{esct0} is the probability of having $k$ traps among the $c$ neighbours of a given trap that lie lower, i.e.\ have a greater depth, and $c-k$ traps that lie higher. Calling the depth of the given central trap $E$, this gives
\begin{align}
  a_k &= \binom{c}{k} \int\d E \, \rho_E(E) P(\ldots, E_k > E, E_{k+1}< E,\ldots) 
  \\
      &= \binom{c}{k} \int\d E \, \rho_E(E) \bigg(\int_{E}^\infty \d E' \rho_E(E') \bigg)^k \times \notag 
      \\
      &\times \bigg(\int_0^{E} \, \d E' \rho_E(E') \bigg)^{c-k} 
      \label{beta_integral}
      \\
      &= \binom{c}{k}  \frac{(c - k)! k!}{(c+1)!} 
      = \frac{1}{c+1} \, ,
\end{align}
where in the initial integrand we assumed a specific ordering of the lower and higher neighbours and compensated for this by 
the binomial coefficient prefactor. The integral in~\eqref{beta_integral} with the variable change $q=\int_0^E \d E'\rho_E(E')$ evaluates to a Beta function as used in the line below.

\section{Scaling of persistence function}
\label{perscal}

We derive here the large $t$-scaling of the persistence function, which from Eq.~\eqref{persinum} is: 
\begin{align}
P(0, t)   &= \int \d E  \mathrm{e}^{-E} \left( \int \d E_1 \, \mathrm{e}^{-E_1} {\rm{e}}^{-(t/c)/(1 + \exp(-\beta(E_1- E) )} \right)^c   .
\label{persinum_app}
\end{align}
For $t/c \gg 1$, the denominator $1+{\rm{e}}^{-\beta(E_1 - E)}$ in the exponent must be large to get a significant contribution and can therefore be approximated by ${\rm{e}}^{-\beta(E_1 - E)}$. The central trap depth $E$ then appears in the combination $(t/c)\mathrm{e}^{-\beta E}$. This suggests a change of integration variable to $\omega=(t/c)^T\mathrm{e}^{-E}$. With a similar transformation $q= \mathrm{e}^{-E_1}/\omega$ for $E_1$, Eq.~\eqref{persinum_app} becomes
\begin{align}
  P(0,t) &= \left(\frac{c}{t} \right)^T \int_0^{(t/c)^T} \d \omega \left[ \omega\int_0^{1/\omega} \d q\,  {\rm{e}}^{-1/(c/t+q^\beta)} \right]^c
 \label{intomega}
  \end{align}
This is still exact but can be simplified for large times, specifically $t/c\gg 1$. The upper boundary in the outer integral can then be replaced by $\infty$ and the $c/t$ in the integrand can be neglected, giving the asymptotic scaling
\beq 
P(0,t) \simeq \pi_c\, t^{-T}
\eeq
The prefactor 
\beq 
\pi_c = c^T \int_0^\infty \d \omega \left(\omega\int_0^{1/\omega} \d q\,  {\rm{e}}^{-q^{-\beta}} \right)^c
\label{prefactor}
\eeq
is, at fixed $T$, just a function of the connectivity $c$. To understand its scaling for large $c$, note that the inner integral is taken to the power $c$ and so contributes only in the small $\omega$-region where it is $1-O(1/c)$. We therefore write it as
\beq 
1-\omega \int_0^{1/\omega} \d q\,  \left(1-{\rm{e}}^{-q^{-\beta}}\right)
\eeq
For small $\omega$ the upper integration boundary can again be taken to $\infty$, giving $1-d_T \omega$ up to higher order corrections, with 
\beq 
d_T = \int_0^\infty \d q\,  \left(1-{\rm{e}}^{-q^{-\beta}}\right)=\Gamma(1-T)
\eeq
where $\Gamma(\cdot)$ is the Euler Gamma function. The prefactor~\eqref{prefactor} of the persistence for large $c$ is then
\beq 
\pi_c = c^T \int_0^\infty \d\omega (1-d_T\omega)^c = c^T \int_0^\infty \d\omega\,\mathrm{e}^{-c d_T\omega} = d_T^{-1}
c^{T-1}
\label{prefactor_large}
\eeq
giving overall for large $c$ and large times $t\geq c$
\beq 
P(0,t) \simeq d_T^{-1}c^{T-1}t^{-T}
\label{persistence_scaling_app}
\eeq
which is the scaling announced in~\eqref{persistence_scaling} in the main text.

We note that the arguments above can be extended to understand the entire large $c$-scaling of the persistence around the entropic-energetic crossover, where $\tilde t = t/c$ is of order unity. Anticipating that again small $\omega$ will dominate, we rescale $\omega=\tilde\omega /c$ in~\eqref{intomega} to get
\begin{align}
  P(0,t) &= \frac{\tilde{t}^{-T}}{c} \int_0^{c\tilde{t}^T} \d \tilde\omega \left\{ 1-\frac{\tilde\omega}{c}\int_0^{c/\tilde\omega} \d q\,  \left[1-{\rm{e}}^{-1/(\tilde{t}^{-1}+q^\beta)}
  \right] \right\}^c
 \label{intomega2}
  \end{align}
For large $c$ the upper integration boundaries again tend to $\infty$, while the integrand $\{\ldots\}^c$ becomes an exponential, leading to
\begin{align}
  {P}(0,t) &= \frac{\tilde{t}^{-T}}{c} \int_0^\infty \d \tilde\omega \exp\left( -\tilde\omega\int_0^\infty \d q\,  \left[1-{\rm{e}}^{-1/(\tilde{t}^{-1}+q^\beta)}
  \right] \right)\\
  &= \frac{\tilde{t}^{-T}}{c} \left( \int_0^\infty \d q\,  \left[1-{\rm{e}}^{-1/(\tilde{t}^{-1}+q^\beta)}
    \right] \right)^{-1} \, .
    \label{rescaled}
\end{align}
This shows that for large $c$, $c{P}(0,t)$ does indeed become a function only of the time $\tilde{t}$ scaled to the crossover time $t^*\sim c$, so it is convenient to introduce the scaled persistence
\beq
\tilde{P}(0,\tilde{t}) \equiv c{P}(0,\tilde{t}/c) \, .
\label{tildeP_definition}
\eeq
For large $\tilde{t}$, Eq.~\eqref{rescaled} directly retrieves the scaling~\eqref{persistence_scaling_app}. For small $\tilde{t}$, on the other hand, the exponential can be linearized so that the $q$-integral becomes
\beq 
\int_0^\infty \d q\,  \frac{1}
{\tilde{t}^{-1}+q^\beta} = \frac{\pi T}{\sin(\pi T)}\,\tilde{t}^{1-T}
\eeq 
and hence
\beq 
\tilde{P}(0,\tilde{t}) = \frac{\sin(\pi T)}{\pi T}\, \frac{1}{\tilde{t}} \, ,
\label{rescaled-persi}
\eeq
The unscaled persistence is then
\beq
P(0, t) = \frac{\tilde{P}(0,t/c)}{c}= \frac{\sin(\pi T)}{\pi T}\, \frac{1}{t}
\eeq
This exhibits the expected $1/t$ decay in the entropically dominated regime, where well before the crossover the connectivity $c$ is irrelevant as long as it is large enough.

\section{Scaling of escape rate distribution for large $c$}
\label{largec}
We show in this appendix what the scaling~\eqref{tildeP_definition} of the persistence function 
\beq 
P(0,t) = c^{-1}\tilde{P}(0,\tilde{t}), \qquad \tilde{t}=t/c
\eeq
for large $c$ implies for the escape rate distribution $\rho_\Gamma(\Gamma)$. Using the relation~\eqref{persis_rho_Gamma}, the rescaled persistence can be written as
\beq 
\tilde{P}(0,\tilde{t}) = c \int \d\Gamma \rho_\Gamma(\Gamma)\mathrm{e}^{-\Gamma c\tilde{t}}
=\int \d\tilde\Gamma \rho_\Gamma(\tilde\Gamma/c)\mathrm{e}^{-\tilde\Gamma\tilde{t}}
\label{large_c_Laplace}
\eeq
where $\tilde\Gamma=c\Gamma$. For this to have a limit for large $c$ requires that also the rescaled relaxation spectrum
\beq
 \tilde{\rho}(\tilde{\Gamma}) =  {\rho}_\Gamma(\tilde{\Gamma}/c)
\label{lap}
\eeq
must become independent of $c$.
This master curve for the large $c$-relaxation rate spectrum 
can expressed in terms of an infinite series using the following steps, starting from~\eqref{rescaled}. First, transform $q \rightarrow \tilde{q} = q \tilde{t}^T$. Second, introduce $u = 1/(1+\tilde{q}^\beta) $. This leads to
\beq
\tilde{P}(0, \tilde{t}) = \left(T \int_0^1 \d u \, (1-u)^{T-1} u^{-T-1} (1- {\rm{e}}^{-\tilde{t}{u}} ) \right)^{-1} \, .
\eeq
We now extract the dominant large $\tilde{t}$-term by decomposing the integral into three parts:
\begin{align}
\tilde{P}(0, \tilde{t})^{-1} &= T \int_{0}^\infty \d u \, u^{-T-1} (1-{\rm{e}}^{-\tilde{t} u} ) \notag \\ &- T \int_0^1 \d u \, u^{-T-1} [1 - (1-u)^{T-1} ] (1 - {\rm{e}}^{-\tilde{t} u} ) \notag \\ &- T \int_1^\infty \d u \, u^{-T-1}(1 - {\rm{e}}^{-\tilde{t} u} ) \, .
\end{align}
The first integral can be computed analytically and gives a pure power law as intended, while the last two terms can be combined into a single integral: \beq
\tilde{P}(0, \tilde{t})^{-1} = \Gamma(1-T) \tilde{t}^T + \int_0^\infty \d u  \, \pi(u) (1- {\rm{e}}^{-\tilde{t} u} ) \, ,
\eeq
where $\Gamma(\cdot)$ is the Euler $\Gamma$-function and $\pi(u)$ is defined as
\beq
\pi(u) = -Tu^{-T-1}
\begin{cases}
  1 - (1 - u)^{T-1} \quad &u <1 \\
  1 \qquad &u \geq 1 \, .
\end{cases}
\eeq
\begin{figure}
  \centering
  \includegraphics[width =  \textwidth]{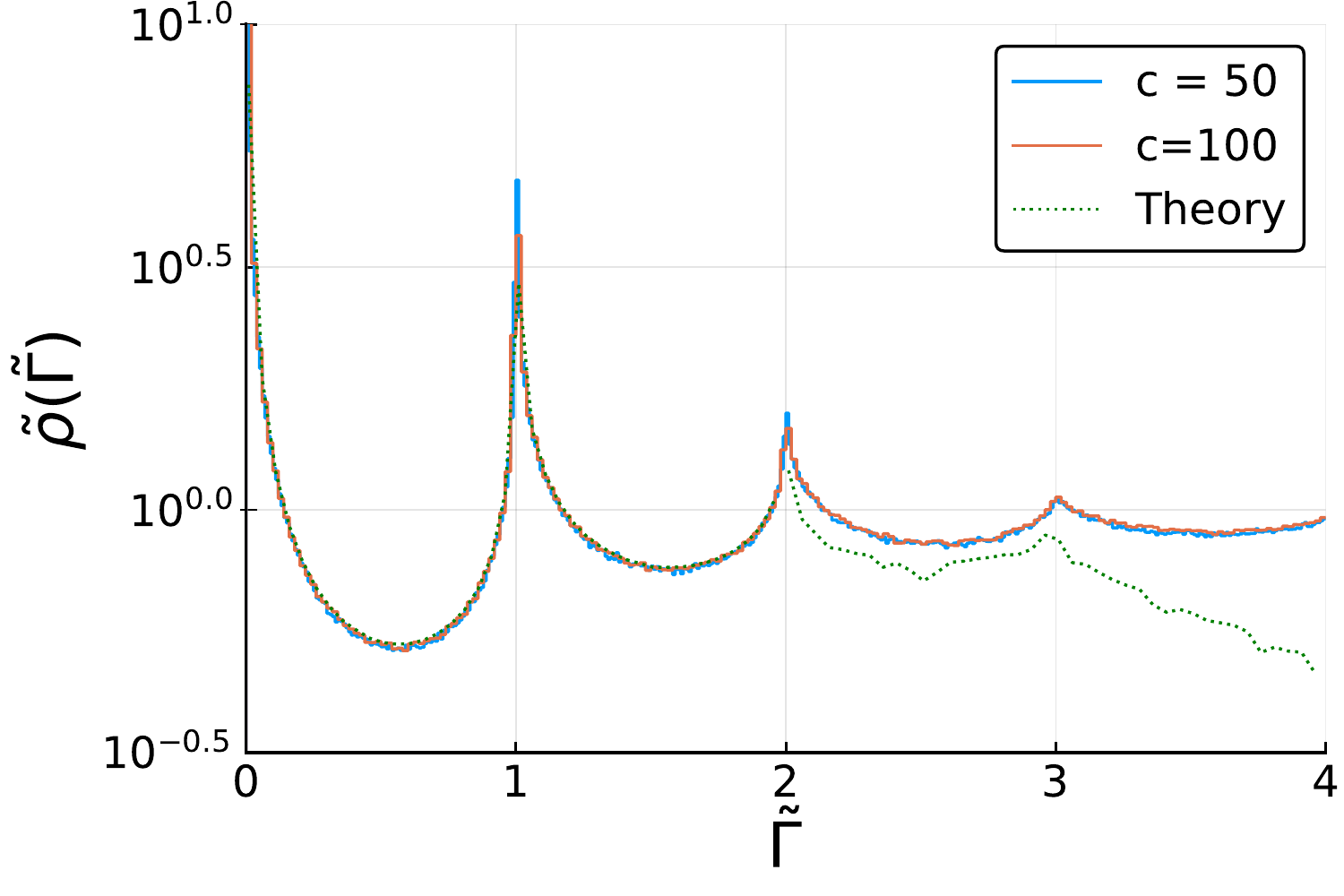}
    \caption{Relaxation rate spectrum plotted against the scaled rate  $\tilde\Gamma = c \Gamma$ for $T =0.2$. Numerical data for two finite $c$ are shown against the theoretical master curve for $c\to\infty$, 
    as predicted by equation~\eqref{predi}. The series was truncated beyond $n=3$, which makes it inaccurate for  $\tilde\Gamma\geq 2$ (dotted line).
    }
\label{figesc}
  \end{figure}
One can check that $\int_0^\infty \d u \pi(u)=0$, giving the further simplification
\begin{align}
  \tilde{P}(0, \tilde{t})^{-1} 
  &=  \Gamma(1-T) \tilde{t}^T \left(1- \frac{\tilde{t}^{-T}}{  \Gamma(1-T) }  \int_0^\infty \d u \, \pi(u) {\rm{e}}^{-\tilde{t} u}\right) \, ,
\end{align}
For large $\tilde{t}$, the inverse 
can now be expanded into a geometric 
series
\begin{align}
  \tilde{P}(0, \tilde{t}) &= \frac{\tilde{t}^{-T}}{\Gamma(1-T) } \left(1- \frac{\tilde{t}^{-T}}{  \Gamma(1-T) }  \int_0^\infty \d u \, \pi(u) {\rm{e}}^{-\tilde{t} u}  \right)^{-1}  \\
                           &= \frac{\tilde{t}^{-T}}{\Gamma(1-T) } \left(1 + \sum_{n \geq 1} \frac{\tilde{t}^{-nT}}{\Gamma^n(1-T) } \int_0^\infty \, \d u\, \pi^{*n}(u) {\rm{e}}^{-\tilde{t} u} \right)
                             \label{finp}
\end{align}
where the superscript ${*n}$ denotes the $n$-th convolution. This expression can now be conveniently inverse Laplace transformed (see (\ref{large_c_Laplace},\ref{lap})) to get the scaled relaxation rate spectrum 
\begin{align}
  \tilde{\rho}(\tilde{\Gamma}) = \frac{\tilde{\Gamma}^{T-1}}{\Gamma(T) \Gamma(1-T)} + \sum_{n \geq 1} \dfrac{\int_0^{\tilde{\Gamma}} \d u \, \pi^{*n}(u)   (\tilde{\Gamma} - u)^{(n+1)T - 1} }{\Gamma((n+1) T) \Gamma^{n+1}(1-T) } \, .
  \label{predi}
\end{align}
The first few terms of this series are straightforward to evaluate numerically. In Figure~\ref{figesc} we compare the resulting prediction for the master curve with numerical data for two different (large) connectivities, finding very good agreement.
The integral associated with $n = 1$ in~\eqref{predi} controls the behaviour at $\tilde{\Gamma} = 1$; it is explicitly given by
  \beq
\int_0^1 \d u \, u^{-T-1}  (1 - (1 -u))^{T-1} (\tilde{\Gamma} - u)^{2T-1} 
\eeq
It is remarkable that this diverges for $T < 1/3$, which is exactly what we found in the case $c = 2$ for the corresponding peak at $\Gamma = 1/2$ (see Eq.~\eqref{cequal2}). This suggests that the structure of the first nonzero peak ($\tilde\Gamma=1$) is independent of $c$, and the same may be true for the peaks at $\tilde\Gamma=2,3,\ldots$ -- of course only for large enough $c$ as $\tilde\Gamma\leq c$ generally.

\section{Cavity equations: Mean field and zero temperature limit }

In this section we show that the results for the mean field limit~\eqref{MF_plateau} and zero temperature (the analogue of~\eqref{esct0}) can be obtained via the cavity equations~(\ref{marginals},\ref{cavities}).

We start by  rewriting the equations in a way that makes them simpler to analyse. 
We start by dividing both sides of~\eqref{cavities} by the factor ${\rm{e}}^{\beta E_k}c$ and correspondingly define rescaled cavity precisions $\tilde{\omega}_k^{(j)} =\omega_k^{(j)} {\rm{e}}^{-\beta E_k}/c$:
\begin{align}
  \tilde{\omega}_k^{(j)}
  & = i (\lambda - i \epsilon) + \sum_{l \in \partial k \setminus j} \frac{i [K(E_k,E_l) \mathrm{e}^{-\beta E_k}/c]\,  \tilde\omega_l^{(k)} }{i [K(E_k, E_l)\mathrm{e}^{-\beta E_l}/c] + \tilde\omega_l^{(k)}}
 \end{align}
From~\eqref{fbeta_def} the combinations in square brackets just give transition rates:
\beq
\tilde{\omega}_k^{(j)}
= i(\lambda - i \epsilon) + \sum_{l \in \partial k \setminus j} \frac{i W_{lk} \tilde{\omega}_l^{(k)} }{i W_{kl}  + \tilde{\omega}_l^{(k)} } \label{scaledcavity}
\eeq
The equations for the scaled marginal precisions follow in the same way from (\ref{marginals}), giving
\beq
\tilde{\omega}_j
= i(\lambda - i \epsilon) + \sum_{k \in \partial j} \frac{i W_{kj} \tilde{\omega}_k^{(j)} }{i W_{jk}  + \tilde{\omega}_k^{(j)} }
\label{scaledmargi}
\eeq
 Bearing in mind that the transition rates scale as $1/c$, the sum in (\ref{scaledcavity}) and hence the typical cavity precision is $O(1)$. The transition rates in the 
 denominators 
 of (\ref{scaledcavity},\ref{scaledmargi}) can thus be neglected for large $c$ and 
\eqref{scaledmargi} simplifies to
\begin{align}
\tilde{\omega}_j  &= i (\lambda - i \epsilon) +  \sum_{k \in \partial j} {i W_{kj} } \, , \\
&= i (\lambda - i \epsilon) +  i \Gamma_{j}  \, .
\label{omegatilde_equals_Gamma}
\end{align}
The spectral density becomes (cf. \ Eqns~\eqref{rhosingle},~\eqref{rholambda}):
\begin{align}
  \rho(\lambda) &= \lim_{\epsilon \rightarrow 0} \frac{1}{\pi N} \sum_{j=1}^N {\rm{Re}} (1/\tilde{\omega}_j) 
  \label{rho_scaled}
  \, \\
                &= \lim_{\epsilon \rightarrow 0} \frac{1}{\pi N} \sum_{j=1}^N  {\rm{Re}} \left( \frac{1}{i(\lambda + \Gamma_j) + \epsilon } \right) \\
                  &= \langle \delta(\lambda + \Gamma) \rangle \, .
\end{align}
Thus in the mean field limit the distribution of escape rates becomes equal to the distribution of relaxation rates.

For the case $T = 0$, on the other hand, the cavity equation~\eqref{scaledmargi} becomes
\begin{align}
  \tilde{\omega}_j = i(\lambda - i \epsilon) + \frac{1}{c}\sum_{k \in \partial j} i {\Theta(E_k-E_j)} \, ,
\end{align}
giving for the spectral density
\begin{align}
  \rho(\lambda)  &= \lim_{\epsilon \rightarrow 0} \frac{1}{\pi N} \sum_{j=1}^N  {\rm{Re}} \left( \frac{1}{i[\lambda + \sum_{k \in \partial j} \Theta(E_k-E_j)/c] + \epsilon } \right) \\
  &= \left \langle \delta \left( \lambda + \sum_{k \in \partial j} \frac{\Theta(E_k-E_j)}{c}  \right) \right \rangle \\
  &= \frac{1}{c+1} \sum_{k=0}^c \delta \left( \lambda + \frac{k}{c} \right) \, ,
\end{align}
where in the last line we have used the result from Appendix~\ref{prefa}.

\section{Cavity equations:  Small $\lambda$-limit and large $c$-limit }
\label{cavlarge}

In this section we will obtain the small $\lambda$ limit~\eqref{small2} for the spectral density from the cavity equations~(\ref{marginals},\ref{cavities}). We will do this first for finite $c$ and then show how the large $c$-behaviour of the prefactor~\eqref{small} can be extracted. 
We start by writing~\eqref{scaledcavity} explicitly as
\begin{align}
\tilde{\omega}_k^{(j)}
&=  i(\lambda - i \epsilon) + i {\rm{e}}^{-\beta E_k} \sum_{l \in \partial k \setminus j} \frac{ {\rm{e}}^{\beta E_l} \tilde{\omega}_l^{(k)} }{i  + c \tilde{\omega}_l^{(k)} (1 +{\rm{e}}^{-\beta(E_k - E_l)})}  \, .
\label{otildes}
\end{align}
In the population picture that one obtains for $N\to\infty$, the analogous relation for the marginal precisions reads
\begin{align}
  \tilde{\Omega}_c = i (\lambda - i \epsilon)  +  i {\rm{e}}^{-\beta E} \sum_{k=1}^c \frac{ {\rm{e}}^{\beta E_k} \tilde{\omega}_k }{i    + c \tilde{\omega}_k  (1 +{\rm{e}}^{-\beta(E - E_k)})}
  \label{tildeO}
\end{align}
and these marginal precisions feed into the spectral density~\eqref{rho_scaled}, whose population form is
\begin{equation}
  \rho(\lambda) = \lim_{\epsilon \rightarrow 0} \frac{1}{\pi} {\rm{Re}} \left \langle \frac{1}{\tilde{\Omega}_c(\{\tilde{\omega}_l, E_l\}, E)} 
  \right \rangle_{(\{\tilde{\omega}_l, E_l \}, E) } \, ,
  \label{rhorescaled}
\end{equation}

Now in the limit of small $\lambda$ we expect the solution of the cavity equations to produce cavity precisions $\tilde{\omega}_k^{(l)}$ that are  purely imaginary, up to a real part of  $O(\epsilon)$~\cite{margiotta2018spectral}.
This allows us to simplify the  expression~\eqref{rhorescaled} for the spectral density as follows:
\begin{align}
  &\rho(\lambda) =  \\
  &= \lim_{\epsilon \rightarrow 0} \frac{1}{\pi} {\rm{Re}}  \left \langle \frac{1}{ O(\epsilon)   +  i \left(\lambda + {\rm{e}}^{-\beta E} \sum_{k=1}^c \frac{ {\rm{e}}^{\beta E_k} \tilde{\omega}_k }{i    + c \tilde{\omega}_k  (1 +{\rm{e}}^{-\beta(E - E_k)})} \right)} 
    \right \rangle \\
  &= \left \langle {\delta\left(\lambda + {\rm{e}}^{-\beta E} \sum_{k=1}^c \frac{ {\rm{e}}^{\beta E_k} \tilde{\omega}_k }{i    + c \tilde{\omega}_k  (1 +{\rm{e}}^{-\beta(E - E_k)})} \right)} 
    \right \rangle \, .
    \label{rhoappr}
\end{align}
For small $\lambda$ one sees that contributions to the $\delta$-function come from large $E$, which makes sense as slow relaxation rates should be associated with activation from the deepest traps in the landscape (with $E - \ln c \gg 1$) that are typically surrounded by higher neighbours. In this regime we can drop the exponential from the denominator in~\eqref{rhoappr} to get
\begin{align}
  \label{rh1}
\rho(\lambda) &\approx \left \langle {\delta\left(\lambda + {\rm{e}}^{-\beta E} \sum_{k=1}^c \frac{ {\rm{e}}^{\beta E_k} \tilde{\omega}_k }{i    + c \tilde{\omega}_k } \right)}  
\right \rangle 
\\
              &=    \int \d E \int \d \xi_c \rho_E(E)\rho_\xi(\xi_c) \delta(\lambda + {\rm{e}}^{-\beta E} \xi_c)
                \label{rh2}
\end{align}
with
\beq
\xi_{c}  =  \sum_{k=1}^{c} \dfrac{ {\rm{e}}^{\beta E_k} \tilde{\omega}_k }{i    + c \tilde{\omega}_k  } \,.
\label{xi}
\eeq
With this, the original average in~\eqref{rh1} over $E$ and the pairs $(\omega_k, \E_k)$ has been translated into the average over $E$ and the effective variable $\xi_c$, which has an $E$-independent distribution $\rho_\xi(\xi_c)$. By evaluating the integral over the exponential energy distribution $\rho_E(E)$ we then arrive at
\beq
   \rho(\lambda) = \kappa_c |\lambda|^{T-1},
\qquad 
\kappa_c = T
   \int_0^\infty \d \xi_c \rho_\xi(\xi_c) \xi_c^{-T} \, .    \label{kappa_c_result}
\eeq

Up to here we have accomplished our first aim, i.e.\ to derive the power law dependence~\eqref{small2} of the spectral density for small $\lambda$ from the cavity equations. The prefactor $\kappa_c$ still has to be found numerically from the $\lambda\to0$ cavity equations, but we can obtain its scaling for large $c$ analytically.
Taking the limit of large $c$ in \eqref{xi}, the $i$ in the denominator can be neglected and one has
\beq 
\xi \approx \frac{1}{c}
\sum_{k=1}^c \mathrm{e}^{\beta E_k}
\label{xi_appr}
\eeq
Now the $\tau_k=\mathrm{e}^{\beta E_k}$ with $E_k$ drawn from $\rho_E(E)=\mathrm{e}^{-E}$ have a distribution with a power law tail $\sim \tau^{-T-1}$ and hence a divergent mean for $T<1$. 
The sum in (\ref{xi_appr}) is therefore dominated by its largest term~\cite{bouchaud1990anomalous} for which $E_k \approx \ln c$, giving
\beq 
\xi \sim c^{-1}\mathrm{e}^{\beta \ln c} = c^{\beta-1}
\label{xi_scaling}
\eeq
This shows that 
\beq 
\kappa_c \sim \xi^{-T} \sim c^{-T(\beta-1)}=c^{T-1}
\eeq
which is the result (\ref{small}) announced in the main text.

The above scaling argument rests on simplifying~\eqref{xi} using the approximation that $c\tilde\omega_k/i\gg 1$ for typical cavity precisions $\tilde\omega_k$. In fact we had shown in (\ref{omegatilde_equals_Gamma}) that (for $\lambda\to 0$, and ignoring the $O(\epsilon)$ real part) the cavity precisions $\tilde\omega_k=i\Gamma(E_k)$ are the escape rates, which scale as $\sim \mathrm{e}^{-E_k}$. So for the deepest traps with $E_k\approx \ln c$, $c\tilde\omega_k/i$ is just of order unity and such traps therefore make a contribution to (\ref{xi}) that is somewhat smaller than we estimated. However, one can nonetheless show that these traps and even rarer, deeper ones still make a contribution to $\xi$ that scales as (\ref{xi_scaling}).

\section{Single Defect Approximation}
\label{sda}

We derive here the small $|\lambda|$-scaling of the spectral density, using a single defect approximation to the solution of the cavity equations, as used for the Bouchaud model in~\cite{margiotta2019glassy}. The baseline for the approximation is the solution of the cavity equations~(\ref{selfc},\ref{omegas}) in the limit $T \rightarrow \infty$, where the dynamics becomes that of a random walk. At $\beta = 0$, $K(E_k, E_l)$ becomes $1/2$ and the r.h.s.~of the cavity equation~\eqref{omegas} is
\beq
\Omega_{c-1}
(\{\omega_l\}) = i \lambda_\epsilon c + \sum_{l=1}^{c-1} \frac{i \omega_l}{i + 2\omega_l} \, .
\label{sdaomegas}
\eeq
In this scenario the energetic disorder (trap depths) no longer plays any role and all nodes are equivalent. Accordingly it turns out that the distribution of cavity precisions becomes a delta distribution peaked at the value $\bar{\omega}$ that solves~\eqref{sdaomegas}, i.e.\ 
\beq
\bar{\omega} = i \lambda_\epsilon c + \frac{i (c-1) \bar{\omega}}{i + 2\bar{\omega}} \, .
\eeq
which gives \begin{align}
  \bar{\omega} &= \frac{1}{4} \bigg( i[c - 2 + 2c(\lambda- i \epsilon)] \\
  &+ \sqrt{-4 + 4c - c^2 - 4c^2(\lambda - i \epsilon) - 4c^2(\lambda- i\epsilon)^2} \bigg) \, .\notag
\end{align}
{The joint distribution of cavity precisions and trap depths is then $\zeta(\omega, E) = \delta(\omega - \bar{\omega})\rho_E(E)$.}
We will be interested in the limit of small $\lambda$, where one can expand $\bar{\omega}$ as
\beq
\bar{\omega} \approx i\frac{(c-2)}{2} + \mathcal{O}(\lambda - i\epsilon) \, ,
\label{barap}
\eeq
The next step in the approximation is to consider that a single node of the $T\to\infty$ network is substituted by a ``defect'' that feels the actual $T$, thus  making its own trap depth and that of its neighbours relevant for the calculation. Thus for a finite $T$ the spectral density is computed as (cf.\ Equation~\eqref{rholambda})
\beq
\rho^A(\lambda) = \lim_{\epsilon \rightarrow 0} \frac{1}{\pi} {\rm{Re}} \left \langle \frac{ {\rm{e}}^{\beta E} c}{\Omega_c(\bar{\omega}, \{E_l\}, E) } \right \rangle_{ \{E_l\},E} \, .
\label{rhoa}
\eeq
We now insert the approximation~\eqref{barap} for $\bar{\omega}$ into $\Omega_c$:
\begin{align}
  \Omega_c(\bar{\omega}, \{E_l\}, E) \approx i (\lambda - i\epsilon) {\rm{e}}^{\beta E} c + \sum_{l=1}^c \frac{i K(E, E_l) (c - 2)}{2 K(E, E_l) + c - 2}
\end{align}
so that~\eqref{rhoa} becomes
\beq
\rho^A(\lambda) \approx \lim_{\epsilon \rightarrow 0} \frac{1}{\pi} {\rm{Re}}    \left \langle \frac{ {\rm{e}}^{\beta E} c }{\epsilon {\rm{e}}^{\beta E} c + i \tilde{K}(E, \{E_l\})} \right \rangle_{ \{E_l\},E}
\label{rhoA_small_lambda}
\eeq
with
\beq
\tilde{K}(E, \{E_l\}) =  \lambda {\rm{e}}^{\beta E} c + \sum_{l=1}^c \frac{K(E, E_l) (c - 2)}{2 K(E, E_l) + c - 2} \, .
\label{f}
\eeq
Taking the limit $\epsilon \rightarrow 0$ in~\eqref{rhoA_small_lambda} gives, just like in the original resolvent trick~\eqref{specific},
\begin{align}
\rho^A(\lambda) &\approx \left \langle \delta 
\left(
\frac{\tilde{K}(E, \{E_l\})}{{\rm{e}}^{\beta E}c}
\right ) \right \rangle_{ \{E_l\},E} 
\\
&= \left \langle \delta \left( \lambda + \frac{1}{c} \sum_{l=1}^c \left( 1 + \frac{2 {\rm{e}}^{\beta E}}{c-2} + \frac{{\rm{e}}^{\beta E}}{{\rm{e}}^{\beta E_l}} \right)^{-1} \right) \right \rangle_{ \{E_l\},E} \, .
\end{align}
This means that if we introduce the variables
\beq
y_l =  \left( 1 + \frac{2 {\rm{e}}^{\beta E}}{c-2} + \frac{{\rm{e}}^{\beta E}}{{\rm{e}}^{\beta E_l}}  \right)^{-1} \, , \quad Y = \frac{1}{c} \sum_{l=1}^c y_l \, ,
\label{yl}
\eeq
then the spectral density can be expressed as
\beq
\rho^A(\lambda) \approx \left \langle \delta( \lambda + Y) \right \rangle_{Y, E} = \int \d Y \int \d E \, \delta(\lambda + Y) p(Y| E) \rho_E (E) \, .
\label{speca}
\eeq
It thus remains for us to find the distribution of $Y$ given $E$, $p(Y|E)$. The simplest estimator is obtained for large $c$, where $Y$ consists of the sum of a large number of terms and so can be replaced by its mean, i.e. $Y \approx \langle y_l \rangle$:
\beq
Y \approx \int_0^\infty \d E_1 \, \rho_E(E_1)  \left( 1 + \frac{2  {\rm{e}}^{\beta E}}{c-2} + \frac{{\rm{e}}^{\beta E}}{{\rm{e}}^{\beta E_1}}  \right)^{-1} \, .
\eeq
This integral yields a hypergeometric function that in the limit of $E \gg 1$ can  be approximated as
\beq
Y = \frac{\pi T}{\sin(\pi T)} {\rm{e}}^{-E} \left(1 + \frac{2{\rm{e}}^{E/T}}{c-2} \right)^{T-1}  =: g(E) \, .
\label{ge}
\eeq
With this estimate the conditional probability in~\eqref{speca} is $P(Y| E) = \delta(Y - g(E))$ so that our single defect approximation reduces to
\beq
\rho^{A}(\lambda) = \frac{\rho_E(E)}{|g'(E)|} \bigg \lvert_{E = g^{-1}(|\lambda|)} \, ,
\label{aaa}
\eeq
Evaluation of this formula requires the inversion of~\eqref{ge}, which cannot be done in closed form. However, bearing in mind that for small $\lambda$ the corresponding $E$ will be large we can approximate further
\beq
1 + \frac{2{\rm{e}}^{E/T}}{c-2}   \approx  \frac{2{\rm{e}}^{E/T}}{c-2} \, .
\eeq
With this~\eqref{aaa} can be evaluated explicitly as
\beq
\rho^{A}(\lambda) = T |\lambda|^{T-1}   \left(\frac{c-2}{2} \right)^{T(T-1)} \left(\frac{\sin(\pi T)}{\pi T} \right)^T \, .
\eeq
As discussed in the main text, this gives the correct scaling with $\lambda$ in the small $\lambda$-regime that we have considered in this appendix, while the exponent for the dependence on $c$ is off by a factor $T$.  One can check by direct numerical sampling of $Y$ (data not shown) that this discrepancy arises not from our approximate evaluation of $P(Y|E)$ but from the single defect approximation itself, i.e.\ from neglecting temperature effects in nodes that are not direct neighbours of the chosen central node.

\end{document}